\documentclass[amsmath,amssymb,floatfix,aps,superscriptaddress,nofootinbib,notitlepage,pra]{revtex4-2}

\usepackage{amsmath,amsthm,amssymb,bm,graphicx,float,xcolor,changes,mathtools,enumitem}
\usepackage[colorlinks=true,urlcolor=purple,linkcolor=blue,citecolor=green,bookmarks=false]{hyperref}
\allowdisplaybreaks

\newcommand{\be}{\begin{equation}}
\newcommand{\ee}{\end{equation}}
\newcommand{\6}{\partial}
\newcommand{\inti}{\int_{-\infty}^{+\infty}}
\newcommand{\bj}{\boldsymbol{j}}
\newcommand{\bl}{\boldsymbol{\lambda}}
\newcommand{\ba}{\boldsymbol{\alpha}}
\newcommand{\bq}{\boldsymbol{q}}
\newcommand{\bmu}{\boldsymbol{\mu}}
\newcommand{\la}{\lambda}

\begin{document}

\title{Nonequilibrium dynamics in one-dimensional strongly interacting two-component gases}

\author{Ovidiu I. P\^{a}\c{t}u}
\affiliation{Institute for Space Sciences, Bucharest-Măgurele, R 077125, Romania}

\begin{abstract}
The derivation of determinant representations for the space-, time-, and temperature-dependent correlation functions of the
impenetrable Gaudin-Yang model  in the presence of a trapping potential is presented. These representations are valid in both
equilibrium and nonequilibrium scenarios like the ones initiated by a sudden change of the
confinement potential. In the equal-time case our results are shown to be equivalent to a multicomponent generalization of
Lenard's formula from which Painlev\'e transcendent representations for the correlators can be obtained in the case of harmonic
trapping and Dirichlet and Neumann boundary conditions. For a system in the quantum Newton's cradle setup the determinant
representations allow for an exact numerical investigation of the dynamics and even hydrodynamization  which is outside the
reach of Generalized Hydrodynamics or other approximate methods. In the case of a sudden change in the trap's frequency,  we
predict a many-body bounce effect, not present in the evolution of the density profile, which causes a nontrivial periodic
narrowing of the momentum distribution with amplitude depending on the statistics of the particles.
\end{abstract}

\maketitle

\section{Introduction}

The study of nonequilibrium phenomena represents one of the most active area of research in modern physics. Due to the unprecedented
degree of control over interactions, dimensionality and statistics the field of ultracold gases represents the ideal testing ground
for various nonequilibrium scenarios in which isolated many-body systems can be accurately observed \cite{GBL13,CCGO11,MVBF21}.
One-dimensional (1D) systems are of  particular interest as they can realize integrable systems which are experimentally accessible
and  in which analytical results can verify and complement more general approximate methods. The realization that integrable and
near-integrable models in 1D do not thermalize \cite{KWW06,RDO08}, as it was shown in the pioneering quantum Newton's cradle experiment
\cite{KWW06}, reignited  interest in the study of such systems resulting in the introduction of powerful techniques like the Quench
Action \cite{CE13,C16} and Generalized Hydrodynamics \cite{CDY16,BCNF16} (GHD). While initial investigations focused on single component
systems in recent years several studies on multicomponent systems, which present a richer phenomenology like spin-charge separation
and the spin-incoherent regime, have also appeared in the literature \cite{IN17,MBPC18,SMS18,ZVR19,WYCZ20,SCP21,TCB21,SCP22,ASYP21,P22,P23}.

In  1D continuum systems with infinitely repulsive contact interactions (also known as the Tonks-Girardeau regime) the correlation
functions can be computed as determinants opening the way for the exact investigation of the dynamics \cite{KBI}. For periodic boundary
conditions and no external potential  determinant representations were obtained in Refs. \cite{L66,KS90,PKA08} for the single component
Lieb-Liniger (LL) model  and in Refs. \cite{BL87,IP98,P19} for the two-component Gaudin-Yang (GY) model. The general case of systems
in external trapping potentials  has been addressed only recently  for the bosonic LL model in \cite{SGPM21} and generalized for arbitrary
statistics (anyons) in \cite{SGPM21,W22,P22b} (for equal-time correlators similar representations results were derived earlier in the
case of harmonic trapping in \cite{FFGW03b,P03,MPC16,H16} and in rather general nonequilibrium scenarios in \cite{PB07,delC08,AGBK17b,P20}).
In this article we derive determinant representations for the space-, time-, and temperature-dependent correlation functions of the
arbitrary statistics Gaudin-Yang model in the presence of a trapping potential which can also depend on time. Our results are valid
in both equilibrium and nonequilibrium scenarios which can be realized in current experiments. In the equal-time case we show the equivalence
of the determinant representation with a multicomponent generalization of Lenard's formula \cite{L66}. Lenard's formula makes transparent the
connection between the correlation functions of the GY model and the gap probabilities in certain random matrix ensembles which were previously
calculated \cite{FFGW03}. In this way Painlev\'e transcendent representations for the correlators of finite size systems at zero temperature
can be easily derived. We also employ our results for the investigation of the dynamics in two experimentally relevant nonequilibrium scenarios:
the sudden change in the trap's frequency and the quantum Newton's cradle setup.  In the first scenario we discover a collective many-body
bounce effect similar with the one described and investigated by Atas {\it et al.} \cite{ABGK17} in the case of single component systems (see
also \cite{P20}). The effect can be seen in the periodic narrowing of the  momentum distribution function (MDF) when the gas is maximally
compressed and is not present in noninteracting systems subjected to the same quench. The amplitude of the narrowing is dependent on the
statistics. Very recently, the phenomenon of hydrodynamization \cite{FHS18}, which describes the rapid onset of hydrodynamics on the fastest
available scale  in a system quenched with an energy much larger than its ground state energy, has been observed in the LL model \cite{LZGR23}.
The main feature  of hydrodynamization in the QNC setup, the rapid change of energy in the momentum modes between the Bragg peaks, cannot be
captured by GHD but it can be accurately monitored using our determinant representation. We perform a detailed investigation
of hydrodynamization in the GY model  highlighting the differences between the two-component and single component systems.

The plan of the paper is as follows. In Sec.~\ref{s1} we introduce the Gaudin-Yang model, the eigenstates, wavefunctions and the correlators.
In Sec.~\ref{s2} we present results for the form factors and in Sec.~\ref{s3} the determinant representations for the correlators. The
particular case of equal-time correlators and the equivalence with Lenard's formula is described in Secs.~\ref{smain} and \ref{s5}. The dynamics
in the case of variable frequency  can be found in Sec.~\ref{s6} and the investigation of hydrodynamization is presented in Sec.~\ref{sqnc}.
We conclude in Sec.~\ref{s8}. Technical details regarding the derivation of the determinant representations, the equivalence with Lenard's
formula and the thermodynamics of the trapped GY model can be found in several Appendices.

\section{The anyonic Gaudin-Yang model in the presence of an external potential}\label{s1}

We consider a one-dimensional system of particles  with two internal states and infinite repulsive contact interactions in the presence of
an external confining potential which can also depend on time. In second quantization the Hamiltonian can be written as
\begin{align}\label{ham}
H=\int \, dx\, \frac{\hbar^2}{2m}\6_x\boldsymbol{\Psi}^\dagger\6_x\boldsymbol{\Psi}+ g:(\boldsymbol{\Psi}^\dagger\boldsymbol{\Psi})^2:
+(V(x,t)-\mu) \boldsymbol{\Psi}^\dagger\boldsymbol{\Psi}+B(\boldsymbol{\Psi}^\dagger\sigma_z\boldsymbol{\Psi})\, ,
\end{align}
where $m$ is the mass of the particles, $g=\infty$ characterizes the strength of the interaction, $ \boldsymbol{\Psi}=(\Psi_\uparrow(x),
\Psi_\downarrow(x))^T$, $ \boldsymbol{\Psi}^\dagger=(\Psi_\uparrow^\dagger(x),\Psi_\downarrow^\dagger(x))$ and $:\ \  :$ denotes normal
ordering. In (\ref{ham}) $\mu$ is the chemical potential, $B$ the magnetic field, $\sigma_z$ the third Pauli matrix and
$\Psi_{\uparrow,\downarrow}(x)$ are anyonic fields satisfying the following commutation relations $(\alpha,\beta\in\{\uparrow,\downarrow\}):$
\begin{subequations}\label{comm}
\begin{align}
\Psi_\alpha(x)\Psi_{\beta}^\dagger(y)&=-e^{-i\pi\kappa\, \mbox{\small{sign}}(x-y)}\Psi_\beta^\dagger(y)\Psi_\alpha(x)+\delta_{\alpha,\beta}
\delta(x-y)\, ,\\
\Psi_\alpha(x)\Psi_{\beta}(y)&=-e^{i\pi\kappa\, \mbox{\small{sign}}(x-y)}\Psi_\beta(y)\Psi_\alpha(x)\, ,\\
\Psi_\alpha^\dagger(x)\Psi_{\beta}^\dagger(y)&=-e^{i\pi\kappa\, \mbox{\small{sign}}(x-y)}\Psi_\beta(y)\Psi_\alpha(x)\, ,
\end{align}
\end{subequations}
with $\mbox{sign}(x)=x/|x|$, $\mbox{sign}(0)=0$ and $\kappa\in [0,1]$ is the statistics parameter (note that an equally valid choice for the
statistics parameter could have been $\kappa\in[-1,0]$). For $x\ne y$, as we vary the statistical parameter the commutation relations
(\ref{comm}) interpolate continuously between the ones for two-component fermions at $\kappa=0$ and two-component bosons at $\kappa=1$. At
coinciding points $x=y$ the commutation relations (\ref{comm}) are fermionic in nature and, therefore, double occupancy, even of particles
of opposite spin, is excluded. We will call (\ref{ham}) the anyonic Gaudin-Yang model as it represents the natural generalization to fractional
statistics of the fermionic and bosonic models introduced and studied first by Gaudin \cite{G67} and Yang \cite{Y67}. The study of anyonic
systems in 1D is now a mature field with important results derived both in single component (like the anyonic Lieb-Liniger model)
\cite{K99,G06,BGO06,BG06,AN07,PKA07,CM07,CS09,HZC09,KLMR11,HC12,WRDK14,AFS16,PC17,SPC20,PSC20,HK20,BJEP21,MD21,ZNIG22} and multicomponent
systems \cite{OAE00,BFGLZ08,YCLF12,SPK12,Z16,CGS16,P19}.

We will consider both static and time-dependent external potentials. In the static case we will consider trapping potentials of the
type $V(x)=a_\nu |x|^\nu$ , $\nu\ge 1$ (the usual harmonic trapping is obtained for $\nu=2$ and $a_\nu=m\omega_0^2/2)$ but our results are
also valid in the case of more general trapping potentials or systems with  Dirichlet or Neumann boundary conditions in a box of dimension
$L$ (in the Dirichlet case the potential can be thought as $V(x)=0$ for $x\in[0,L]$ and $V(x)=\infty$ for  $x\notin[0,L]$). The case
without external potential $V(x)=0$ and periodic boundary conditions was investigated in \cite{IP98,P19}.
In the time-dependent case we will consider potentials which present a sudden change at $t=0$ resulting in quantum quenches but as we will
see our results are valid also in other nonequilibrium scenarios like the quantum Newton's cradle setup \cite{KWW06}. Prototypical examples are
the change of  the trap's frequency of a harmonic trapping potential
\begin{align}
V(x,t)=\left\{
\begin{array}{ll}
m\omega_0^2x^2/2\, , &t\le 0\, ,\\
m\omega_1^2x^2/2\, , &t> 0\, ,
\end{array}
\right.
\ \ \ \ \omega_0\ne \omega_1\, ,
\end{align}
(the free expansion after the release from a trap is a particular case with $\omega_1=0$) and the change of the shape of the trap
\begin{align}
V(x,t)=\left\{
\begin{array}{ll}
a_{\nu}|x|^\nu\, , &t\le 0\, ,\\
a_{\nu'}|x|^{\nu'}\, , &t> 0\, ,
\end{array}
\right.
\ \ \ \ \nu\ne \nu'\, .
\end{align}
In the time-dependent case we have an initial Hamiltonian denoted by $H_I$ and a final Hamiltonian denoted by $H_F$. In order to compute the
time evolution of the correlators we will consider either the equilibrium groundstate or a thermal state of the initial Hamiltonian $H_I$ but
the subsequent evolution will be given by the final Hamiltonian $H_F$. Of course, in the static case we have $H_I=H_F$.

In order to highlight the differences between the two-component and single component systems we will make frequent comparisons with results
for the anyonic Lieb-Liniger (LL) model \cite{K99,BGO06,AN07,PKA07}  described by the Hamiltonian
\begin{align}\label{hamLL}
H_{LL}=\int \, dx\, \frac{\hbar^2}{2m}\6_x\Psi^\dagger\6_x\Psi+ g \Psi^\dagger\Psi^\dagger\Psi\Psi
+(V(x,t)-\mu) \Psi^\dagger\Psi\, ,
\end{align}
where now $\Psi^\dagger(x)$ and $\Psi(x)$ are single component anyonic fields satisfying similar commutation relations like (\ref{comm}).
For $\kappa=1$ the Hamiltonian (\ref{hamLL}) reduces to the usual bosonic LL model while for $\kappa=0$ it describes free fermions (single
component fermions do not ``feel" the contact interaction).
From now on we will consider $\hbar=k_B=1$ with $k_B$ the Boltzmann constant.

\subsection{Eigenstates at $t=0$}

While the homogeneous fermionic and bosonic GY models are integrable for any value of the repulsive interaction (the proof of the integrability
in the anyonic case is an open problem) the addition of an external potential breaks this integrability with the exception of zero
and infinite repulsion. In the impenetrable case, which is the focus of this paper,  we will introduce in a constructive fashion a complete
set of eigenstates of the initial Hamiltonian  which solve the many-body Schr\"odinger equation, satisfy the hard-core condition and have
the proper symmetry when exchanging two particles of the same type.

At $t=0$ the eigenstates of the initial Hamiltonian for a system of $N$ particles of which $M$ have spin down are given by
\begin{align}\label{eigen}
|\Phi_{N,M}(\bj,\bl)\rangle=\int\prod_{k=1}^N dx_k\, \sum_{\alpha_1,\cdots,\alpha_N=\{\uparrow,\downarrow\}}^{[N,M]}
\chi_{N,M}^{\alpha_1\cdots\alpha_N}(\boldsymbol{x}|\bj,\bl)\Psi_{\alpha_N}^\dagger(x_N)\cdots\Psi_{\alpha_1}^\dagger(x_1)|0\rangle\, ,
\end{align}
where $\boldsymbol{x}=(x_1,\cdots,x_N)$, the $[N,M]$ over the sum sign means that we sum over combinations of $\alpha$'s such that $M$ of
them  are spin down and  $N-M$ are spin up and $|0\rangle$ is the Fock vacuum satisfying $\Psi_\alpha(x)|0\rangle=\langle 0|\Psi_\alpha^
\dagger(x)=0$ for all $\alpha$ and $x$. The eigenstates (\ref{eigen}) are indexed by two sets of unequal numbers $\bj=(j_1,\cdots,j_N)$ and
$\bl=(\lambda_1,\cdots,\lambda_M)$ (their meaning will be made clear below) and the normalized wavefunctions are $[\ba=(\alpha_1\cdots
\alpha_N)]$
\begin{align}\label{wavef}
\chi_{N,M}^{\ba}(\boldsymbol{x}|\bj,\bl)=\frac{1}{N!N^{M/2}}\left[\sum_{P\in S_N}\theta(x_{P(1)}<\cdots< x_{P(N)})e^{i\frac{\pi\kappa}{2}
\sum_{1\le a<b\le N}\, \mbox{\small{sign}}(x_a-x_b)}\eta_{N,M}^{(\ba,P\ba)}(\bl)\right] \det_N\left[\phi_{j_a}(x_b)\right]\, ,
\end{align}
with the sum being taken over all the permutations of $N$ elements denoted by $S_N$, $\theta(x_{1}<\cdots< x_{N})=\prod_{l=2}^N\theta(x_l-
x_{l-1})$  with $\theta(x)$ the Heaviside function and $P\ba=(\alpha_{P(1)}\cdots\alpha_{P(N)})$. In the right hand side of (\ref{wavef}) the
Slater  determinant is constructed from the eigenfunctions of the initial single-particle Hamiltonian defined by
\begin{align}\label{hamsp}
H_I^{SP}(x)\phi_j(x)=\varepsilon(j)\phi_j(x)\, ,\ \ \
H_I^{SP}(x)=-\frac{1}{2m}\frac{\6^2}{\6x^2}+V(x,t\le 0)\, .
\end{align}
For example, if $V(x,t\le 0)=m\omega_0^2 x^2/2$ then $\phi_j(x)$ is the $j$-th Hermite function of frequency $\omega_0$ and $\varepsilon(j)=
\omega_0(j+1/2)$. The spin sector is described by $\eta_{N,M}^{(\ba,P\ba)}(\bl)$ which are the wavefunctions of the $XX$ spin-chain on a lattice
with $N$ sites and $M$ spins down. Explicitly, we have \cite{CIKT93}
\begin{align}\label{wavefs}
\eta_{N,M}^{(\ba,P\ba)}(\bl)=\prod_{1\le a<b\le M}\mbox{sign}(n_b-n_a)\det_M \left[e^{ i n_a \lambda_b}\right]\, ,
\end{align}
where $\bl=(\lambda_1,\cdots,\lambda_M)$ are solutions  of the Bethe ansatz equations for the spin problem
\be
e^{i \lambda_a N}=(-1)^{M-1}\, ,\ \ a=1,\cdots,M\, ,
\ee
and $\boldsymbol{n}=(n_1,\cdots,n_M)$ are a set of integers which are the positions of the spin down particles in the set $(\alpha_{P(1)},\cdots,
\alpha_{P(N)})$. For example, if $\ba=(\downarrow\downarrow\uparrow\uparrow)$ and $P=(3214)$ then the set of $n$'s for $\ba$ is $\boldsymbol{n}=
(1,2)$  while for $P\ba=(\uparrow\downarrow\downarrow\uparrow)$ we have $\boldsymbol{n}=(2,3)$.

The wavefunctions (\ref{wavef}) are the natural generalization of the Bethe Ansatz solution for the impenetrable Gaudin-Yang model \cite{IP98}
in the presence of an external potential. They exhibit factorization of the spin and charge degrees of freedom characteristic of impenetrable
multicomponent systems \cite{OS90,EFGKK,DFBB08,GCWM09,VFJV14,LMBP15,YC16,DBBR17,YAP22}, solve the many-body Schr\"odinger equation, vanish when
two coordinates coincide (hard-core condition), have the appropriate symmetry when exchanging two particles of the same type
\be\label{gensym}
\chi_{N,M}^{\alpha_1\cdots \alpha_i,\alpha_{i+1}\cdots\alpha_N}(x_1,\cdots,x_i,x_{i+1},\cdots,x_N)=-e^{i\pi\kappa\, \mbox{\small{sign}}(x_i-x_{i+1})}
\chi_{N,M}^{\alpha_1\cdots \alpha_{i+1},\alpha_{i}\cdots\alpha_N}(x_1,\cdots,x_{i+1},x_{i},\cdots,x_N)\, ,
\ee
and form a complete set. We should point out that while we chose for the description of the spin sector the $XX$ spin chain wavefunctions an equally
valid alternative, but not as computationally efficient, is represented by the  $XXX$ spin chain wavefunctions.

The eigenstates (\ref{eigen}) are normalized
$\langle\Phi_{N,M}(\bj,\bl)|\Phi_{N',M'}(\bj',\bl')\rangle=\delta_{N,N'}\delta_{M,M'}\delta_{\bj,\bj'}\delta_{\bl,\bl'}$ and satisfy
$H_I |\Phi_{N,M}(\bj,\bl)\rangle=E_{N,M}(\bj)|\Phi_{N,M}(\bj,\bl)\rangle$ with
\be\label{energy}
E_{N,M}(\bj)=\sum_{l=1}^N(\varepsilon(j_l)-\mu+B)-2MB\, ,
\ee
The spectrum of the impenetrable anyonic Gaudin-Yang model is independent on the spin state and statistics resulting in large degeneracies of the
groundstate and excited states. At zero temperature even an infinitesimal magnetic field totally polarizes the system which is then equivalent  to
the LL model.

\subsection{Time evolution of the eigenstates}

The time evolution of the eigenstates (\ref{eigen}) can be easily determined by taking into account that due to the impenetrability of the particles
the spin degrees of freedom are effectively frozen which means that the dynamics is encoded in the charge degrees of freedom (the proof is almost
identical with the one presented in  \cite{P23} for the Bose-Fermi mixture). This means that the time evolved eigenstates are
\begin{align}\label{i1}
e^{- i t H_F}|\Phi_{N,M}(\bj,\bl)\rangle=e^{-it[(-\mu+B)N-2BM]}|\Phi_{N,M}(t|\bj,\bl)\rangle\, ,
\end{align}
where $|\Phi_{N,M}(t|\bj,\bl)\rangle$ is described by (\ref{eigen}) with the time dependent wavefunction given by
\begin{align}\label{waveft}
\chi_{N,M}^{\ba}(\boldsymbol{x},t|\bj,\bl)=\frac{1}{N!N^{M/2}}\left[\sum_{P\in S_N}\theta(x_{P(1)}<\cdots< x_{P(N)})e^{i\frac{\pi\kappa}{2}
\sum_{1\le a<b\le N}\, \mbox{\small{sign}}(x_a-x_b)}\eta_{N,M}^{(\ba,P\ba)}(\bl)\right] \det_N\left[\phi_{j_a}(x_b,t)\right]\, .
\end{align}
In the time independent case ($H_I=H_F$) the time evolved single particle orbitals appearing in the Slater determinant on the right hand side of
(\ref{waveft}) are given by
\begin{align}
\phi_j(x,t)=e^{-i \varepsilon(j) t}\phi_j(x,0)\, ,
\end{align}
with $\phi_j(x,0)$ and $\varepsilon(j)$ the eigenfunctions and eigenenergies of the single particle Hamiltonian (\ref{hamsp}).
In the time dependent case ($H_I\ne H_F$) $\phi_j(x,t)$ is the unique solution of the Schr\"odinger equation
\begin{align}
i\frac{\6 \phi_j(x,t)}{\6 t}=H_F^{SP}(x)\phi_j(x,t)\, ,\ \
H_F^{SP}(x)=-\frac{1}{2m}\frac{\6^2}{\6x^2}+V(x,t> 0)\, ,
\end{align}
satisfying the initial boundary condition $\phi_j(x,0)=\phi_j(x)$ where $\phi_j(x)$ is an eigenfunction of the initial single particle Hamiltonian
(\ref{hamsp}).

\subsection{Correlators}

We are interested in deriving efficient numerical representations for the space-, time-, and temperature-dependent correlation functions of the
Gaudin-Yang model for a system prepared in a grandcanonical thermal state of the initial Hamiltonian $H_I$ described by the chemical potential $\mu$,
magnetic field $B$ and temperature $T$.  We will investigate two correlators $(\sigma\in\{\uparrow,\downarrow\})$:
\begin{align}\label{defgm}
g_{\sigma}^{(-)}(x,t;y,t')&=\langle\Psi_\sigma^\dagger(x,t)\Psi_\sigma(y,t')\rangle_{\mu,B,T}\, ,\nonumber\\
&=\mbox{Tr}\left[e^{-H_I/T} \Psi_\sigma^\dagger(x,t)\Psi_\sigma(y,t')\right]/\mbox{Tr}\left[e^{-H_I/T}\right]\, ,\nonumber\\
&=\sum_{N=0}^\infty\sum_{M=0}^{N+1} \sum_{j_1<\cdots<j_{N+1}}\sum_{\lambda_1<\cdots<\lambda_M}\frac{e^{-E_{N+1,M}(\bj)/T}}{\mathcal{Z}}
\langle\Phi_{N+1,M}(\bj,\bl)|\Psi_\sigma^\dagger(x,t)\Phi_\sigma(y,t')|\Phi_{N+1,M}(\bj,\bl)\rangle\, ,
\end{align}
and
\begin{align}\label{defgp}
g_{\sigma}^{(+)}(x,t;y,t')&=\langle\Psi_\sigma(x,t)\Psi_\sigma^\dagger(y,t')\rangle_{\mu,B,T}\, ,\nonumber\\
&=\mbox{Tr}\left[e^{-H_I/T} \Psi_\sigma(x,t)\Psi_\sigma^\dagger(y,t')\right]/\mbox{Tr}\left[e^{-H_I/T}\right]\, ,\nonumber\\
&=\sum_{N=0}^\infty\sum_{M=0}^N \sum_{q_1<\cdots<q_N}\sum_{\mu_1<\cdots<\mu_M}\frac{e^{-E_{N,M}(\bq)/T}}{\mathcal{Z}}
\langle\Phi_{N,M}(\bq,\bmu)|\Psi_\sigma(x,t)\Phi_\sigma^\dagger(y,t')|\Phi_{N,M}(\bq,\bmu)\rangle\, ,
\end{align}
where
\begin{align}
\mathcal{Z}=\mbox{Tr}\left[e^{-H_I/T}\right] =\sum_{N=0}^\infty\sum_{M=0}^N \sum_{q_1<\cdots<q_N}\sum_{\mu_1<\cdots<\mu_M}e^{-E_{N,M}(\bq)/T}\, ,
\end{align}
is the partition function of the initial Hamiltonian in the grandcanonical ensemble described by $\mu$ and $B$ at temperature $T$. In (\ref{defgm})
and (\ref{defgp}) the time evolution is dictated  by the final Hamiltonian $H_F$ and the evolved operators are given by
\begin{align}
\Psi_\sigma(x,t)=e^{i H_F t} \Psi_\sigma(x) e^{-i H_F t}\, ,\ \ \ \Psi_\sigma^\dagger(x,t)=e^{i H_F t} \Psi_\sigma^\dagger(x) e^{-i H_F t}\, .
\end{align}
The real space densities $\rho_\sigma(x,t)$  and momentum distribution functions (MDFs) $n_\sigma(k,t)$ can be obtained from the equal-time correlator
$g_\sigma^{(-)}(x,t;y,t)$  using
\begin{align}\label{defrn}
\rho_\sigma(x,t)=g_\sigma^{(-)}(x,t;x,t)\, , \ \ \ n_\sigma(k,t)=\frac{1}{2\pi}\int\int e^{- i k(x-y)}g_\sigma^{(-)}(x,t;y,t)\, dxdy\, .
\end{align}

\section{Form factors}\label{s2}

The derivation of the determinant representations for the correlators (\ref{defgm}) and (\ref{defgp}) is relatively involved requiring
several steps. In the first step we are going to compute the form factors which appear in the decomposition of the mean values of bilocal
operators present in the definition of the correlators. Then, the form factors can be summed using a method which can be understood
as a modification of the Cauchy-Binet formula \cite{KS90,KBI,IP98} resulting in a determinant representation for the mean values. In the
third step we take the thermodynamic limit and use the von Koch's determinant formula to obtain the desired result. Before we present
the derivation we make an important observation. Due to the $SU(2)$ symmetry of the Hamiltonian (\ref{ham}) it is sufficient to study only
one type of correlators, the other type can be easily obtained using the relation $ g_\uparrow^{(\pm)}(x,t;y,t'|B)=g_\downarrow^{(\pm)}
(x,t;y,t'|-B)\, . $

We will start by computing the form factors. The mean values of bilocal operators appearing in the right hand-side of (\ref{defgm}) and
(\ref{defgp}) can be  written as sums over form factors as follows.  Using the completeness of the eigenstates
$ \bm{1}=\sum_{N=0}^\infty\sum_{M=0}^N \sum_{\substack{q_1<\cdots<q_N\\  \mu_1<\cdots<\mu_M}}|\Phi_{N,M}(\bq,\bmu)\rangle\langle\Phi_{N,M}
(\bq,\bmu)|\, ,$
we obtain (the bar denotes complex conjugation)
\begin{align}\label{i30}
\langle \Phi_{N+1,M}(\bj,\bl)|\Psi_\sigma^\dagger(x,t)\Psi_\sigma(y,t')|\Phi_{N+1,M}(\bj,\bl)\rangle
&=\sum_{\substack{q_1<\cdots<q_N\\  \mu_1<\cdots<\mu_{\bar{M}}}}
\overline{\mathcal{F}}_{N,M}^{(\sigma)}(\bj,\bl;\bq,\bmu|x,t)
\mathcal{F}_{N,M}^{(\sigma)}(\bj,\bl;\bq,\bmu|y,t')\, ,
\end{align}
and
\begin{align}\label{i31}
\langle \Phi_{N,\bar{M}}(\bq,\bmu)|\Psi_\sigma(x,t)\Psi_\sigma^\dagger(y,t')|\Phi_{N,\bar{M}}(\bq,\bmu)\rangle
&=\sum_{\substack{j_1<\cdots<j_{N+1}\\  \lambda_1<\cdots<\lambda_M}}
\mathcal{F}_{N,M}^{(\sigma)}(\bj,\bl;\bq,\bmu|x,t)
\overline{\mathcal{F}}_{N,M}^{(\sigma)}(\bj,\bl;\bq,\bmu|y,t')\, ,
\end{align}
where
\be\label{formf}
\mathcal{F}_{N,M}^{(\sigma)}(\bj,\bl;\bq,\bmu|x,t)=\langle\Psi_{N,\bar{M}}(\bq,\bmu)|\Psi_\sigma(x,t)|\Phi_{N+1,M}(\bj,\bl)\rangle\, ,
\ee
is a general form factor of the $\Psi_\sigma(x,t)$ operator on arbitrary states $(\bj,\bl)$ in the $(N+1,M)$-sector and $(\bq,\bmu)$ in the
$(N,\bar{M})$-sector ($|\Phi_{N,-1}(\bj,\bl)\rangle=0$ by convention) and
\be
\bar{M}=\left\{\begin{array}{lll} M&\ \ \mbox{ if } &\sigma=\uparrow\, ,\\
M-1&\ \ \mbox{ if } &\sigma=\downarrow\, .
\end{array}\right.
\ee
The form factor of the $\Psi_\sigma^\dagger(x,t)$ operator is given by the complex conjugate of (\ref{formf}) i.e.,
$\overline{\mathcal{F}}_{N,M}^ {(\sigma)}(\bj,\bl;\bq,\bmu|x,t)$. In the following we will not write explicitly the dependence of the form
factors on the state parameters when  there is no risk of confusion.

The derivation of the determinant representation for the form factors is presented in Appendix \ref{aff}. It reads
\begin{align}\label{formff}
\mathcal{F}_{N,M}^{(\sigma)}(\bj,\bl;\bq,\bmu|x,t)&=\frac{e^{i t\mu_\sigma}e^{-i\frac{\pi\kappa N}{2}}}{N^{\bar{M}/2}(N+1)^{M/2}}(-1)^{\delta_{\sigma,\downarrow}(M-1)}\det_M
B_\sigma(\bl,\bmu) \det_{N+1} D(\bj,\bq|x,t)\, ,
\end{align}
where $\mu_\uparrow=\mu-B\, , \mu_\downarrow=\mu+B$. This representation is factorized with the charge degrees of freedom being described
by  $D(\bj,\bq|x,t)$ a square matrix of dimension $N+1$ and elements
\begin{align}
[D(\bj,\bq|x,t)]_{ab}=\left\{
\begin{array}{lll}
f(j_a,q_b|x,t) & \mbox{ for } & a=1,\cdots,N+1\, ; \, b=1,\cdots,N\, ,\\
\phi_{j_a}(x,t) & \mbox { for } & a=1,\cdots,N+1\, ;\,  b=N+1\, .
\end{array}
\right.
\end{align}
with ($L_+$ is the right boundary of the system)
\be
f(j,q|x,t)=\delta_{j,q}-\left(1-e^{i\pi\kappa} \overline{\omega}\nu\right)\int_x^{L_+}\overline{\phi}_q(v,t)\phi_j(v,t)\, dv\, .
\ee
where $\omega=e^{i\Lambda}\, , \nu=e^{i \Theta}$ with  $\Lambda=\sum_{a=1}^M \lambda_a\, ,\ \Theta=\sum_{b=1}^{\bar{M}}\mu_b\, . $
The spin degrees of freedom are described by determinants of matrices with dimension $M$ and elements $[B_\uparrow(\bl,\bmu)]_{ab}=
\sum_{n=1}^N e^{i n(\lambda_a-\mu_b)}\, , a,b=1,\cdots,M$ in the spin-up case and
\begin{align}
[B_\downarrow(\bl,\bmu)]_{ab}=\left\{
\begin{array}{lll}
\sum_{n=1}^N e^{i n(\lambda_a-\mu_b)} & \mbox{ for } & a=1,\cdots,M\, ; \, b=1,\cdots,M-1\, ,\\
1  & \mbox { for } & a=1,\cdots,M\, ;\,  b=M\, ,
\end{array}
\right.
\end{align}
for the spin-down case.

\section{Determinant representations for the correlation functions}\label{s3}

Using the formulas for the form factors from the previous section the mean values (\ref{i30}) and (\ref{i31}) can be summed obtaining
rather cumbersome expressions. The situation becomes simpler in the thermodynamic limit, or, more precisely in the large $N$ limit. The
necessary calculations are presented in Appendix \ref{ad}. Before we present our results we need to introduce certain relevant functions
and parameters. First we introduce the parameter $\gamma=\left(1+e^{2B/T}\right)$ and the  building block of our representations  the
function
\be\label{defft}
f(j,q|\eta,x,t)=\delta_{j,q}-\left[1-e^{i(\pi\kappa-\eta)}\right]\int_x^{\infty}\overline{\phi}_q(v,t)\phi_j(v,t)\, dv\, .
\ee
We will also need
\be\label{deftheta}
F(\gamma,\eta)=1+\sum_{p=1}^\infty \gamma^{-p}\left(e^{i\eta p}+e^{-i \eta p}\right)\, ,
\ \ \ \vartheta(a)=\frac{e^{-B/T}}{2\cosh(B/T)+e^{(\varepsilon(a)-\mu)/T}}\, ,
\ee
and we note that  $ F(\gamma=1,\eta)=2\pi \delta(\eta)\, $  and that $\theta(a)$ can be understood as the Fermi function for the spin up
particles of the two-component system (see Appendix \ref{a1}). Now we can state on the main results of our paper. The space-, time-, and
temperature-dependent correlation functions of the anyonic GY model in a trapping potential have the following determinant representations:
\begin{align}\label{gm}
g_\uparrow^{(-)}(x,t;y,t')=\frac{e^{-i(t-t')\mu_\uparrow}}{2\pi}\int_{-\pi}^\pi F(\gamma,\eta)
\left[\det\left(1+\gamma V^{(T,-)}(\eta)+R^{(T,-)}\right)-\det\left(1+\gamma V^{(T,-)}(\eta)\right)\right]\, d\eta\, ,
\end{align}
with $[V^{(T,-)}]_{ab}=\sqrt{\vartheta(a)}(U_{ab}^{(-)}-\delta_{a,b})\sqrt{\vartheta(b)}$ and $[R^{(T,-)}]_{ab}=\sqrt{\vartheta(a)}
R_{ab}^{(-)}\sqrt{\vartheta(b)}$
where $U^{(-)}$ and $R^{(-)}$ are infinite matrices with elements
\begin{subequations}
\begin{align}
U_{ab}^{(-)}(x,t;y,t'|\eta)&=\sum_{q=1}^\infty \overline{f}(a,q|\eta,x,t) f(b,q|\eta,y,t')\, ,\ \ a,b=1,\cdots\, ,\label{defumT}\\
R_{ab}^{(-)}(x,t;y,t')&=\overline{\phi}_{a}(x,t)\phi_{b}(y,t')\, ,\ \  a,b=1,\cdots\, .\label{defrmT}
\end{align}
\end{subequations}
For the second type of correlators the following representation is valid
\begin{align}\label{gp}
g_\uparrow^{(+)}(x,t;y,t')=\frac{e^{i(t-t')\mu_\uparrow}}{2\pi}\int_{-\pi}^\pi F(\gamma,\eta)
\left[\det\left(1+\gamma V^{(T,+)}(\eta)-\gamma R^{(T,+)}(\eta)\right)+(g-1)\det\left(1+\gamma V^{(T,+)}(\eta)\right)\right]\, d\eta\, ,
\end{align}
with $[V^{(T,+)}]_{ab}=\sqrt{\vartheta(a)}(U_{ab}^{(+)}-\delta_{a,b})\sqrt{\vartheta(b)}$ and $[R^{(T,+)}]_{ab}=\sqrt{\vartheta(a)}
R_{ab}^{(+)}\sqrt{\vartheta(b)}$
where $U^{(+)}$ and $R^{(+)}$ are infinite matrices with elements
\begin{align}
U^{(+)}_{ab}(x,t;y,t'|\eta)&=\sum_{j=1}^\infty f(j,b|\eta,x,t)\overline{f}(j,a|\eta,y,t')\, ,\ \ a,b,=1,\cdots\, ,\\
R^{(+)}_{ab}(x,t;y,t'|\eta)&=\bar{e}_a(x,t;y,t'|\eta)e_b(x,t;y,t'|\eta)\, ,\ \ a,b,=1,\cdots\, ,\\
e_a(x,t;y,t'|\eta)&=\sum_{j=1}^\infty f(j,a|\eta,x,t)\overline{\phi}_j(y,t')\, ,\ \ a=1,\cdots\, ,\\
\bar{e}_a(x,t;y,t'|\eta)&=\sum_{j=1}^\infty\overline{f}(j,a|\eta,y,t')\phi_j(x,t)\, ,\ \ a=1,\cdots\, ,
\end{align}
and
\be
g(x,t;y,t')=\sum_{j=1}^\infty\phi_j(x,t)\overline{\phi}_{j}(y,t')\, .
\ee
We make an observation. The terms in the square parenthesis of (\ref{gm}) and (\ref{gm}) represent the single component equivalent
field-field correlators of LL anyons (see \cite{P22b,W22}) with statistics parameter $\kappa-\eta/\pi$ and Fermi function defined in
(\ref{deftheta}). A similar proposal was made in \cite{GQBZ23} in the case of spin-$\frac{1}{2}$ fermions on the lattice. Our results
show that the  transformation introduced in \cite{GQBZ23} it is valid also in the continuum case and can be extended for arbitrary
statistics.

\section{Equal-time correlators}\label{smain}

In order to study the nonequilibrium dynamics of the GY model in several scenarios of interest, like harmonic trapping with variable
frequency or the quantum Newton's cradle setup, it is sufficient to consider the $g^{(-)}_\uparrow(x,t;y,t)$ correlator from  which
the dynamics of the real space densities and momentum distributions can be computed using (\ref{defrn}). In the equal-time case
$t=t'$ the representation for the $g^{(-)}_\uparrow(x,t;y,t)\equiv g^{(-)}_\uparrow(x,y|\,t)$ correlator simplifies considerably.
Using the fact that the time evolved eigenfunctions are orthonormal
\be\label{ortho}
\int_{L_-}^{L_+}\overline{\phi}_q(v,t)\phi_j(v,t)\, dv=\delta_{j,q}\, ,\ \ \sum_{j=1}^\infty\overline{\phi}_j(w,t)\phi_j(v,t)=
\delta(w-v)\, ,
\ee
in Appendix \ref{a2} we show that in the equal-time case the elements of the $V^{(T,-)}$ matrix simplify to
\be\label{i10}
[V^{(T,-)}]_{ab}=-\left(1-e^{-i\, \mbox{\small{sign}}(y-x)[\pi\kappa-\eta]}\right)\mbox{sign}(y-x)\sqrt{\vartheta(a)\vartheta(b)}
\int_x^y\overline{\phi}_a(v,t)\phi_b(v,t)\, dv\, .
\ee
The dependence on $\eta$ is now simple enough that we can integrate in (\ref{gm}). We will denote the difference of determinants
appearing in (\ref{gm}) by $\Xi$. In the case $x\le y$ we have $\Xi=\sum_{n=0}^\infty \gamma^n(e^{-i\pi\kappa}e^{i\eta}-1)^n A(n)$
where $A(n)$ are coefficients that do not depend on $\eta$. We find
\begin{align}
g^{(-)}_\uparrow(x,y|\,t)&=\int_{-\pi}^\pi\frac{d\eta}{2\pi}\left[1+\sum_{p=1}^\infty \gamma^{-p}\left(e^{i\eta p}+e^{-i \eta p}\right)
\right] \sum_{n=0}^\infty \gamma^n(e^{-i\pi\kappa}e^{i\eta}-1)^n A(n)\, ,\nonumber\\
&=\sum_{n=0}^\infty\left(e^{-i\pi\kappa}-\gamma\right)^n A(n)\, ,
\end{align}
where we have used $(e^{-i\pi\kappa}e^{i\eta}-1)^n=\sum_{k=0}^n C^n_k (-1)^{n-k}e^{i\eta k}e^{-i\pi\kappa k}$ and $\int_{-\pi}^\pi
e^{i\eta k}e^{-i\eta p}/(2\pi)=\delta_{k,p}$. In the $y<x$ case we obtain $g^{(-)}_\uparrow(x,t;y,t)=\sum_{n=0}^\infty\left(
\gamma-e^{i\pi\kappa}\right)^n A(n)$. This means that in the equal-time case  the following representation is valid
\be\label{statictfull}
g_\uparrow^{(-)}(x,y|\,t)=\det\left(1+ v^{(T,-)}+r^{(T,-)}\right)-\det\left(1+ v^{(T,-)}\right)\, ,
\ee
with
\begin{subequations}\label{statict}
\begin{align}
[v^{(T,-)}]_{ab}&=-\left[\gamma-e^{-i\, \pi\kappa \mbox{\small{sign}}(y-x)}\right]\mbox{sign}(y-x)\sqrt{\vartheta(a)\vartheta(b)}
\int_x^y\overline{\phi}_a(v,t)\phi_b(v,t)\, dv\, ,\\
[r^{(T,-)}]_{ab}&=\sqrt{\vartheta(a)}\, \overline{\phi}_{a}(x,t)\phi_{b}(y,t) \sqrt{\vartheta(b)}\, .\label{defrt}
\end{align}
\end{subequations}

At zero magnetic field and zero temperature the parameter $\gamma=2$ and $\vartheta(a)=\frac{1}{2}\theta(\mu-\varepsilon(a))$. The
infinite matrices appearing in (\ref{statict}) are replaced with finite matrices of dimension $N$ with $N$ being the number of energy
levels smaller than $\mu$ and  elements
\begin{subequations}\label{statict0}
\begin{align}
[v^{(0,-)}]_{ab}&=-\frac{1}{2}\left(2-e^{-i\, \pi\kappa \mbox{\small{sign}}(y-x)}\right)\mbox{sign}(y-x)
\int_x^y\overline{\phi}_a(v,t)\phi_b(v,t)\, dv\, ,\ \ a,b=1,\cdots,N\, ,\\
[r^{(0,-)}]_{ab}&=\frac{1}{2}\, \overline{\phi}_{a}(x,t)\phi_{b}(y,t)\, ,\ \ a,b=1,\cdots,N\, .
\end{align}
\end{subequations}
We make an important observation. The zero temperature determinant representation (\ref{statictfull}) with matrices (\ref{statict0})
describes the correlators in the spin-incoherent regime \cite{BL87,B91,CZ04a,CZ04b,CSZ05,M04,FB04,F07} which is obtained by taking first
the limit  of infinite repulsion and then $T\rightarrow 0$. Finding a determinant representation for the impenetrable GY model in the
Tomonaga-Luttinger regime, which is obtained by taking first the limit $T\rightarrow 0$ and then the limit of infinite repulsion, is an
open problem \cite{CSZ05,CZ04a}.

We should point out that the representation (\ref{statictfull})  is extremely efficient from the numerical point of view (the main
computational effort comes from the evaluation of the partial overlaps) allowing for the exact investigation of systems with hundreds
of particles at zero temperature and tens of particles at very high temperatures,
which is more than enough for comparison with current experiments (see, for example, the recent experiment \cite{LZGR23} where $N=32$
atoms per tube at zero temperature). Similar representations for single component systems in the continuum can be found in \cite{PB07,delC08,AGBK17b,P20}.

\section{Lenard's formula}\label{s5}

In Ref.~\cite{L66} Lenard used the Bose-Fermi mapping to derive an expansion of single component bosonic correlators in terms of free
fermionic  correlators which was independent on the statistical ensemble and interparticle potential as long as the hard-core
condition was satisfied. In Appendix \ref{a3} we show that the determinant representation (\ref{statictfull}) for the equal-time
correlators is equivalent to the following multicomponent generalization of Lenard's formula:
\begin{align}\label{lenard}
g_\uparrow^{(-)}(x,y|\,t)&=g_\uparrow^{FF}(x,y|\,t)+\sum_{j=1}^\infty\frac{(-\xi)^j}{j!}\int_x^y\, dx_1\cdots\int_x^y\, dx_j\,
g_\uparrow^{FF}\left(
\begin{array}{cccc}
x & x_1 &\cdots & x_j\\
y & x_1 &\cdots & x_j\\
\end{array}
;t
\right)\, ,
\end{align}
with
\be\label{defxi}
\xi=\left[\gamma-e^{-i\, \pi\kappa \mbox{\small{sign}}(y-x)}\right]\mbox{sign}(y-x)\, ,
\ee
and $g_\uparrow^{FF}(x,y|\,t)=\sum_{a=1}^\infty\vartheta(a)\, \overline{\phi}_a(x,t)\phi_a(y,t)$ which  can be understood as the
field-field correlation function of a system of free fermions with Fermi function $\vartheta(a)$ defined in (\ref{deftheta}). In
(\ref{lenard}) we have used the notation
\be
g_\uparrow^{FF}\left(
\begin{array}{cccc}
x & x_1 &\cdots & x_j\\
y & x_1 &\cdots & x_j\\
\end{array}
;t
\right)
=\left|
\begin{array}{cccc}
g_\uparrow^{FF}(x,y|\,t)  & g_\uparrow^{FF}(x,x_1|\,t) &\cdots &g_\uparrow^{FF}(x,x_j|\,t)\\
g_\uparrow^{FF}(x_1,y|\,t)  & g_\uparrow^{FF}(x_1,x_1|\,t) &\cdots &g_\uparrow^{FF}(x_1,x_j|\,t)\\
\vdots& \vdots &\ddots & \vdots\\
g_\uparrow^{FF}(x_j,y|\,t)  & g_\uparrow^{FF}(x_j,x_1|\,t) &\cdots &g_\uparrow^{FF}(x_j,x_j|\,t)
\end{array}
\right|\, .
\ee
The correlator $g_\downarrow(x,y|t)$ has a similar representation as (\ref{lenard})  with $B$ replaced by $-B$ in the expressions
for $\gamma$ and the Fermi function $\vartheta(a)$.

At zero temperature and zero magnetic field  we have $g_\uparrow^{FF}(x,y|\,t)=\frac{1}{2}\sum_{a=1}^N\, \overline{\phi}_a(x,t)
\phi_a(y,t)$ where $N$ is the number of particles for the balanced system in the ground state. The
generalization of Lenard's formula for the balanced system  takes the form
[$g_\uparrow^{(-)}(x,y|\,t)=g_\downarrow^{(-)}(x,y|\,t)$]
\begin{align}\label{lenard0}
g_\uparrow^{(-)}(x,y|\,t)&=\frac{1}{2}\left[g_\uparrow^{FF,0}(x,y|\,t)+\sum_{j=1}^\infty\frac{(-\xi_0)^j}{j!}\int_x^y\, dx_1\cdots
\int_x^y\, dx_j\,
g_\uparrow^{FF,0}\left(
\begin{array}{cccc}
x & x_1 &\cdots & x_j\\
y & x_1 &\cdots & x_j\\
\end{array}
;t
\right)\right]\, ,
\end{align}
with
\be
\xi_0=\frac{1}{2}\left[2-e^{-i\, \pi\kappa \mbox{\small{sign}}(y-x)}\right]\mbox{sign}(y-x)\, ,\ \
g_\uparrow^{FF,0}(x,y|\,t)=\sum_{a=1}^N\, \overline{\phi}_a(x,t)\phi_a(y,t)\, .
\ee

Lenard's formula (\ref{lenard}) is extremely useful in deriving  short distance expansions for the correlators, from which the
Tan contacts which govern the $C(t)/k^4$ tails of the momentum distributions can be extracted, but it can also be used to obtain
Painlev\'e transcendent representations for finite size systems in equilibrium at zero temperature. Let us show how this can be done.
The main observation is that (\ref{lenard0})  can be understood as the first Fredholm minor of the Fredholm integral operator
$1-\xi_0\, \hat{g}_\uparrow^{FF,0}$ acting on $[x,y]$ and with kernel $g_\uparrow^{FF,0}(\lambda,\mu)$. Using Hurwitz formula
\cite{H14} we find
\be\label{lenardresolvent}
g_\uparrow^{(-)}(x,y)=\frac{1}{2}R_\uparrow^{FF}(x,y)\det\left(1-\xi_0\, \hat{g}_\uparrow^{FF,0}\right)\, ,
\ee
with the resolvent satisfying the integral equation
\be\label{resolvent}
R_\uparrow^{FF}(\lambda,\mu)=g_\uparrow^{FF,0}(\lambda,\mu)+\xi_0\int_x^yg_\uparrow^{FF,0}(\lambda,\nu)R_\uparrow^{FF}(\nu,\mu)\, d\nu\, .
\ee
In the particular case at  when $x$ and $y$ are chosen such that they are symmetrical about the origin, say $[-x,x]$, and using
$\frac{d}{dx}\log\det\left(1-\xi_0\, \hat{g}_\uparrow^{FF}\right)=-2 R_\uparrow^{FF}(x,x)$ we obtain
\be
g_\uparrow^{(-)}(-x,x)=\frac{1}{2}R_\uparrow^{FF}(-x,x)\exp\left(-2\int_0^xR^{FF}_\uparrow(t,t)\, dt\right)\, .
\ee
The importance of the previous formula resides in the fact that the quantities $R^{FF}_\uparrow(-x,x)$ and $R^{FF}_\uparrow(t,t)$ have previously been
calculated in terms of Painlev\'e transcendents as part of studies on gap probabilities for certain random matrix ensembles \cite{TW94,WFC00}.
In the harmonic trapping case the relevant ensemble is the Gaussian Unitary Ensemble and the Painlev\'e transcendent representation can be found
in Proposition 5 of \cite{FFGW03} with the parameter $\xi=\left(1-e^{-i\, \pi\kappa }/2\right).$ It is interesting to note that modulo a $1/2$ factor
the correlators of the finite GY model with harmonic trapping  can be expressed in terms  of the same $P_V$ transcendent as the single component bosonic
system, the only difference being in the boundary conditions. In the case of Dirichlet and Neumann boundary conditions the ensemble of interest is
the Jacobi Unitary Ensemble with $a=b=\pm 1/2$ and the transcendent representation can be found in  Proposition 6 of \cite{FFGW03}.

\section{Dynamics in the case of variable frequency}\label{s6}

Using the results of Sec.~\ref{smain} we can investigate the dynamics of the real space densities and momentum distribution functions
of the Gaudin-Yang model in the experimentally relevant case of a trapping potential with variable frequency. We will focus on the case
of free expansion of the gas and the breathing oscillations initiated by a sudden change in the trap's frequency.

In the case of a harmonic potential with variable frequency $V(x,t)= m\omega^2(t)x^2/2$ with $\omega(t\le 0)=\omega_0$ the single particle
eigenfunctions at $t=0$ are the Hermite functions of frequency $\omega_0$
\be\label{hermf}
\phi_j(x)=\frac{1}{\sqrt{ 2^j j!}}\left(\frac{m \omega_0}{\pi}\right)^{1/4}e^{- \frac{m\omega_0^2 x^2}{2}}H_j(\sqrt{m\omega_0}x)\, ,
\ee
where $H_j(x)$ are the Hermite polynomials. The time evolution of the single particle orbitals is given by the scaling transformation
(\cite{PP70}, Chap. VII of \cite{PZ98}):
\be
\phi_j(x,t)=\frac{1}{\sqrt{b(t)}}\phi_j\left(\frac{x}{b(t)},0\right)\exp\left[i\frac{m x^2}{2}\frac{\dot b}{b}-i \varepsilon(j) \tau(t)
\right]\, ,
\ee
with $b(t)$ the solution of the Ermakov-Pinney equation $\ddot{b}+\omega^2(t)b=\omega_0^2/b^3$ and initial boundary conditions $b(0)=1$,
$\dot{b}(0)=0$, $\varepsilon(j)=\omega_0(j+1/2)$ and $\tau(t)=\int_0^t dt'/b^2(t')$. Due to the fact that the dynamics is encoded only in the
charge degrees of freedom  the time evolution of the correlators is given by \cite{ASYP21,P22,P23}
\be\label{i25}
g_\sigma^{(-)}(x,y|\,t)=\frac{1}{b(t)}g_\sigma^{(-)}\left(\left.\frac{x}{b(t)},\frac{y}{b(t)}\right|\, 0\right)e^{-\frac{i}{b}\frac{\dot{b}}
{\omega_0}\frac{x^2-y^2}{2l_o^2}}\, ,
\ee
with $l_o=\sqrt{1/m\omega_0}$ the harmonic oscillator length. The time evolution of the densities is
\be\label{i25b}
\rho_\sigma(x,t)=\frac{1}{b(t)}\rho_\sigma\left(\left.\frac{x}{b(t)}\right|0\right)\, ,
\ee
and in the case of the momentum distributions we have
\be\label{i26}
n_\sigma(k,t)=\frac{b}{2\pi} \int\int\, dxdy\, g_\sigma^{(-)}(x,y| 0) \exp\left[- i b\left(\frac{\dot{b}}{\omega_0}\frac{x^2-y^2}{2l_o^2}
+k(x-y)\right)\right]\, .
\ee
These results show that in order to study the exact dynamics in the case of a system with variable frequency it is sufficient to compute the
correlators at $t=0$ and then use Eqs.~(\ref{i25}) and (\ref{i26}).

\subsection{Free expansion}

Free expansion is described by $\omega(t\le0)=\omega_0$ and $\omega(t\ge 0)=0$ and is ubiquitous in cold gases experiments allowing
for the investigation of the MDF. In this case the solution of the Ermakov-Pinney equation is $b(t)=\left(1+\omega_0^2 t^2\right)^{1/2}$.
We will consider first the case of balanced systems at zero temperature (note that even an infinitesimal magnetic field will polarize
the system at $T=0$ reducing its study to the single component case).
Employing the stationary phase approximation in (\ref{i26}) analytical results on the total asymptotic momentum distribution can be obtained
showing that it is the same as  the MDF of a system of free fermions in the initial trap. This phenomenon is called dynamical fermionization
\cite{RM05,MG05,delC08,GP08,CDDK19,BHLM15,CGK15,XR17} and in the case of the single component bosonic TG gas was experimentally observed in
\cite{WMLZ20}. In the case of bosonic and fermionic spinor gases with any number of components at zero temperature dynamical fermionization
of the total momentum distribution was derived by an Alam {\it et. al} in \cite{ASYP21}. For the anyonic GY model using the explicit form of
the wavefunctions (\ref{wavef}) and the method of \cite{P23} it can be shown that the asymptotic momentum distributions are given by
\be
n_\sigma(k,t\rightarrow\infty)\sim \frac{1}{2} n_{FF}(k)\, ,\ \
n(k,t\rightarrow\infty)\sim  n_{FF}(k)\, ,
\ee
where $n(k,t)=n_\uparrow(k,t)+n_\downarrow(k,t)$  is the total MDF of the GY model and $n_{FF}(k)$ is the MDF of a similar number of free
fermions $N$ in the original trap [$\phi_j(x)$ are defined in (\ref{hermf})]
\be\label{i26b}
n_{FF}(k)=\frac{1}{2\pi}\int\int e^{- i k(x-y)}g_{FF}^{(-)}(x,y)\, dxdy\, ,\ \ \
g_{FF}^{(-)}(x,y)=\sum_{i=0}^{N-1}\overline{\phi}_i(x)\phi_i(y)\, .
\ee
\begin{figure}[h]
\includegraphics[width=1\linewidth]{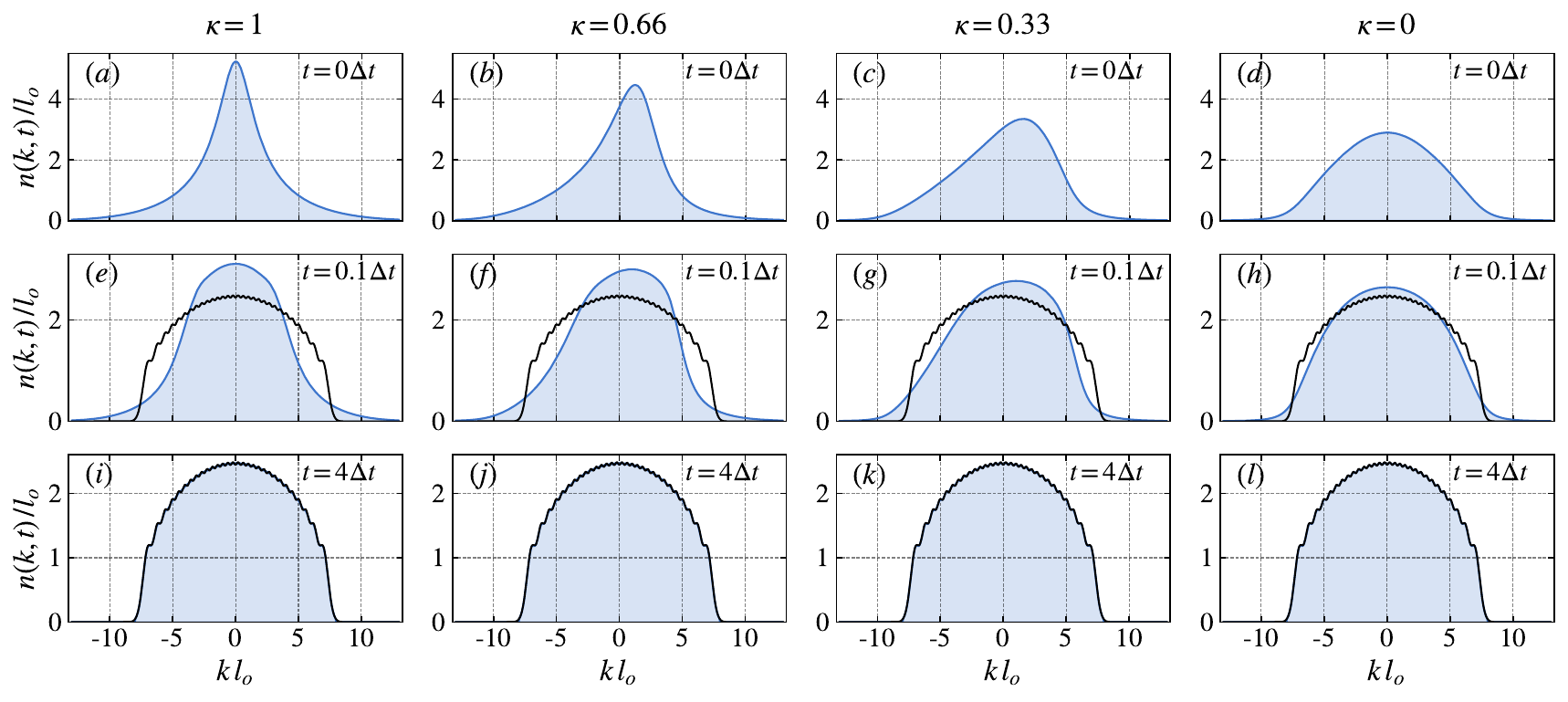}
\caption{Momentum distribution functions before (first row) and after free expansion at  $t=0.1\Delta t$ (second row) and $t=4\Delta t$ (third row)
computed using Eqs.~(\ref{statictfull}), (\ref{statict0}) and (\ref{i26}).
We consider balanced systems of $N=30$ particles at zero temperature ($\omega_0=1, l_o=1, \Delta t=\pi/\omega_0$) and statistics parameter
$\kappa=\{1,0.66,0.33,0\}$. In the second and third row the black line represents  the momentum distribution function of a system of free fermions
with the same number of particles in the initial harmonic trap Eq.~(\ref{i26b}).
}
\label{figexp0}
\end{figure}
\begin{figure}[h]
\includegraphics[width=1\linewidth]{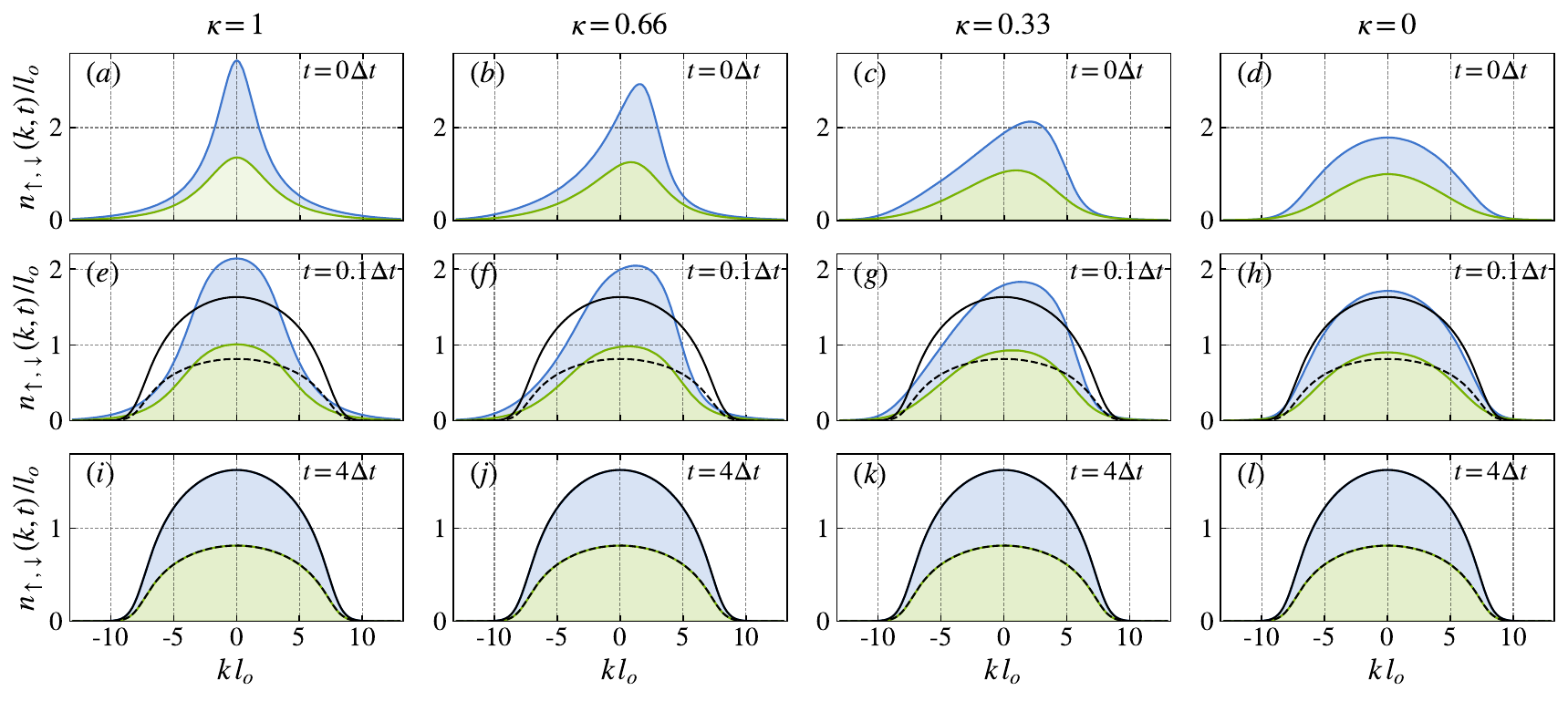}
\caption{Momentum distribution functions of an imbalanced system at finite temperature before (first row) and after free expansion at  $t=0.1\Delta t$
(second row) and $t=4\Delta t$ (third row) computed using Eqs.~(\ref{statictfull}), (\ref{statict}) and (\ref{i26}).
Here we consider systems with $N=30$ particles ($N_\downarrow=20$, $N_\uparrow=10$) at temperature $T=4$
($\omega_0=1, l_o=1,  \Delta t=\pi/\omega_0$) and statistics parameter $\kappa=\{1,0.66,0.33,0\}$. The blue (green) continuous lines represent the MDFs
of the spin-down  (up) particles and in the second and third row the black continuous (dashed) lines represent the analytical result Eq.~(\ref{analytic})
for the spin-down (up) particles.
}
\label{figexpt}
\end{figure}
In Fig.~\ref{figexp0} we present the dynamics of the total MDF for a zero temperature balanced anyonic GY model with $N=30$ particles for
different values of the statistical parameter $\kappa=\{1,\,0.66,\, 0.33,\, 0\}$ and three values of $t$:  before the release from the trap
$t=0$ (first row), immediately after release $t=0.1\pi/\omega_0$ (second row)  and in the asymptotic region $t=4\pi/\omega_0$ (third row). At
$t=0$ the MDF for the bosonic system ($\kappa=1$) presents a visible peak at $k=0$ similar with the one for single component bosons  but  less
pronounced due to the spin-incoherent nature of the system. The fermionic MDF ($\kappa=0$) of the GY model is also smoothened out compared with
the free fermionic MDF which in the presence of the trapping potential presents a number of local maxima equal to the number of particles in the
system. The main feature of the MDF for anyonic systems $(\kappa=\{0.66, 0.33\}$) is the asymmetry which is caused by the broken space invariance
of the commutation relations (\ref{comm}) resulting in $g_\sigma^{(-)}(x,y)=\overline{g_\sigma^{(-)}(y,x)}$ [for the bosonic and fermionic systems
$g_\sigma^{(-)}(x,y)$ is real and we have $g_\sigma^{(-)}(x,y)=g_\sigma^{(-)}(y,x)$]. For all systems at large times after the release from the
trap the asymptotic momentum distribution approaches the symmetric MDF for free fermions in the initial trap (\ref{i26b}) as it can be seen in
the last row of Fig.~\ref{figexp0}.

At finite temperature the situation is more complex. In \cite{P23} it was shown that for a trapped system initially found in a thermal state
described by the chemical potential $\mu$, magnetic field $B$ and temperature $T$ the asymptotic momentum distribution is the same as the one for
spinless free fermions in the initial trap at the same temperature but renormalized chemical potential
\be
\mu'=\mu+T\ln[2\cosh(B/T)]\, .
\ee
Explicitly, the asymptotic MDF for each component reads
\be\label{analytic}
n_\downarrow(k,t\rightarrow\infty)\sim \frac{e^{B/T}}{2\cosh(B/T)}n_{FF}^{\mu'}(k)\, ,\ \
n_\uparrow(k,t\rightarrow\infty)\sim \frac{e^{-B/T}}{2\cosh(B/T)}n_{FF}^{\mu'}(k)\, ,
\ee
and $n(k,t\rightarrow\infty)\equiv n_\downarrow(k,t\rightarrow\infty)+n_\uparrow(k,t\rightarrow\infty)=n_{FF}^{\mu'}(k)$ where $n_{FF}^{\mu'}(k)$
is the MDF of trapped spinless free fermions given by
\be\label{i27}
n_{FF}^{\mu'}(k)=\frac{1}{2\pi}\int\int e^{- i k(x-y)}g_{FF,\mu'}^{(-)}(x,y)\, dxdy\, ,\ \ \
g_{FF,\mu'}^{(-)}(x,y)=\sum_{i=0}^{\infty}\frac{1}{1+e^{(\varepsilon(i)-\mu')/T}}\overline{\phi}_i(x)\phi_i(y)\, .
\ee
In Fig.~\ref{figexpt} we present the time evolution of the MDF after release from the trap for an unbalanced system of $N=30$ particles
($N_\downarrow=20, N_\uparrow=10$) at  temperature $T=4$ which shows the prefect agreement with our analytical predictions for the asymptotic
distributions (\ref{analytic}).

\begin{figure}[ht]
\includegraphics[width=1\linewidth]{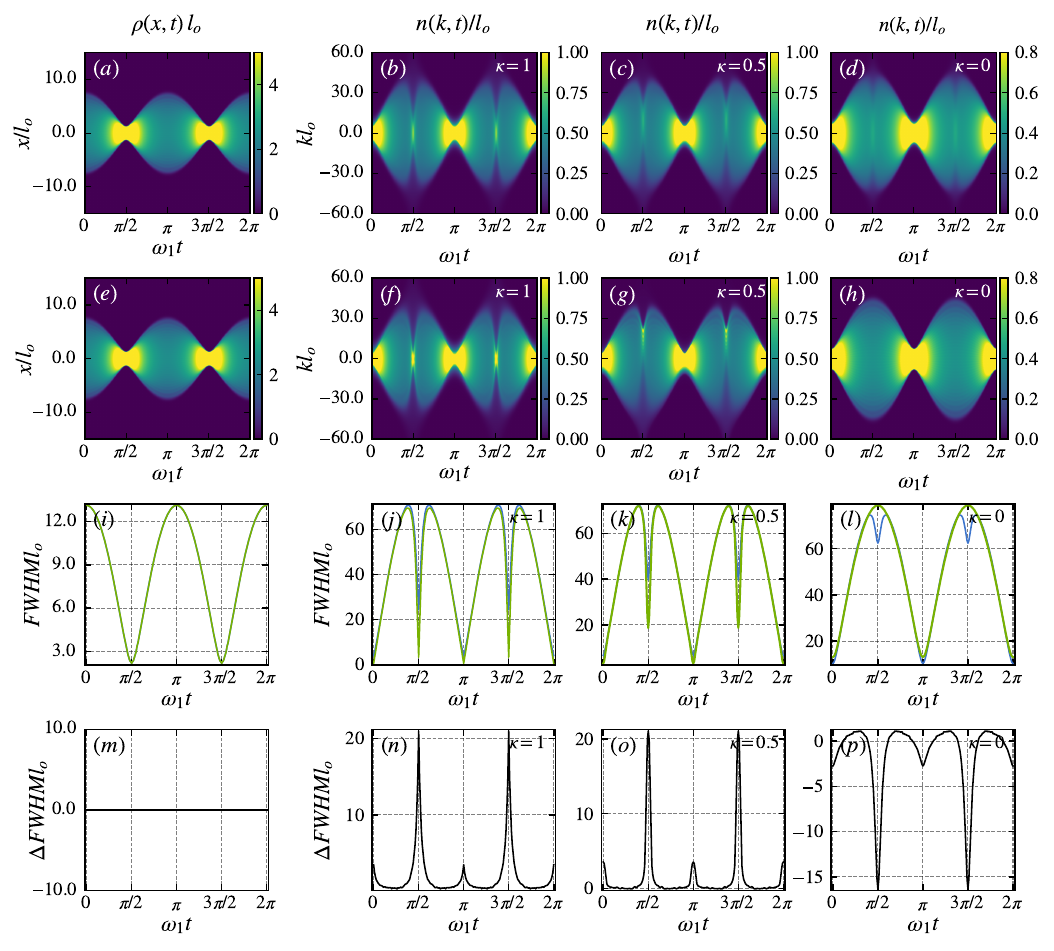}
\caption{Breathing oscillations dynamics in the GY and LL models. First row: a) Time dependence of the total density $\rho(x,t)$ (Eq.~(\ref{i25b})) and total momentum
distribution $n(k,t)$  (Eq.~(\ref{i26b})) for  $\kappa=1$ b), $\kappa=0.5$ c) and $\kappa=0$ d) in the balanced GY model at zero temperature with $N=30$ particles after
a strong quench of the trap  frequency ($\omega_0=1, \omega_1=6\omega_0, \epsilon_0\sim -0.972, l_o=1$). The correlator at $t=0$ is computed with Eqs.~(\ref{statictfull})
and (\ref{statict0}).
Second row: e)-h) Same quantities as above
for the single component  LL model with the same number of particles. Third row: a) Width (FWHM) of the densities [blue (green) line for the GY (LL)
model] and momentum  distributions j)-l). Fourth row: Width difference $\Delta FWHM=FWHM_{GY}-FWHM_{LL}$ for the densities m) and momentum
distributions n)-p).
}
\label{figosc}
\end{figure}

\subsection{Breathing oscillations and collective many-body bounce effect}

A confinement quench in which the trap frequency is suddenly changed to a new value initiates breathing oscillations which can be experimentally
observed \cite{FCJB16,WMLZ20}. We will denote the pre-quench frequency by  $\omega(t\le0)=\omega_0$ and the post-quench frequency by
$\omega(t\ge 0)=\omega_1$. In this case the solution of the Ermakov-Pinney equation is given by $b(t)=\left(1+\epsilon_0\sin^2(\omega_1 t)
\right)^{1/2}$ with $\epsilon_0=\left(\omega_0/\omega_1\right)^2-1$ and describes oscillations between $1$ and $\omega_0/\omega_1$ with period
$\pi/\omega_1$.

In Fig.~\ref{figosc} we present and compare the dynamics of the densities and MDFs for a balanced GY model with $N=30$ particles at zero temperature
and the LL model with the same number of particles subjected to a strong confinement quench $\omega_1=6\omega_0$. The time evolution of the real
space densities, which are the same for both models, is described by  self-similar breathing  cycles $\rho(x,t)=\rho(x/b(t)| 0)/b(t)$ which can
be seen in Fig.~\ref{figosc}a) and Fig.~\ref{figosc}e). The situation is more complex in the case of the MDFs. From Fig.~\ref{figosc}b)-d) we see
that for the GY model and all values of the statistics parameter the MDF dynamics  is no longer self-similar and presents two instances of narrowing:
at $\omega_1t=\pi l, l=0,1,\cdots$ (called outer turning points \cite{ABGK17}) when the real density is the broadest and at $\omega_1 t=\frac{\pi}{2} l,
l=1,2,\cdots$ (called inner turning points) when the gas is maximally compressed.
The additional narrowing at the inner turning point is a manifestation of a many-body collective effect not present in noninteracting systems (see
Fig.~\ref{figosc}h) which can be understood as a self-reflection of the cloud due to the repulsive interactions. In the case of single component bosons
this collective effect was discovered and investigated in  \cite{ABGK17} and in the case of single component anyonic systems in \cite{P20}. The amplitude
of the narrowing at the inner turning points depends on statistics being the largest for bosons ($\kappa=1$) and smallest for fermions ($\kappa=0$).
This can be seen in the evolution of the Full Width at Half Maximum of the MDF presented in Fig.~\ref{figosc}j)-l) and can also be explained in terms
of the repulsive interactions
between the particles: in the bosonic case we have inter and intra-particles interaction while in the fermionic case only particles with opposite pseudo-spins
interact with the anyonic case being in between. We will denote $FWHM_{GY}$($FWHM_{LL}$) the widths of the relevant quantities for the GY (LL) model.
The width differences $\Delta FWHM=FWHM_{GY}-FWHM_{LL}$ plotted in the fourth row of Fig.~\ref{figosc} show that during the time evolution the
largest differences in the MDFs of single and two-component systems occur in the vicinities of the inner and outer turning points. For the bosonic and
anyonic system with $\kappa=0.5$ $\Delta FWHM$ is always positive signalling a broader MDF for the two-component system which is due to the
spin-incoherence of the system (see the discussion in Sec. \ref{sqnc}). In the fermionic case in the vicinities of the inner and outer turning points the
MDF of free fermions is wider than the one for the GY model while in between the inner and outer turning points the opposite is true.

\section{Dynamics in the Newton's quantum cradle setup}\label{sqnc}

In the original Quantum Newton Cradle (QNC) experiment \cite{KWW06} a quasi-1D  ultracold gas of LL bosons in a weakly harmonic trap
is subjected to a sequence of Bragg pulses which splits the initial quasicondensate into two counter-propagating clouds with momenta
centered around $\pm q$. The fact that these clouds undergo repeated oscillations without thermalization like an ordinary gas highlighted
the importance of the large number of conservation laws in the description of nonequilibrium 1D quantum systems. From the theoretical
point of view the dynamics of single component bosons in the QNC setup  has been investigated in \cite{BWEN16,AGBK17b,CDDK19,SBDD19,TDK21}.
Here, we focus on the two-component Gaudin-Yang model (see also \cite{SCP22} for a GHD approach).

First, let us show how our formalism developed in the previous sections can be applied in the QNC setup. Generalizing the results of
\cite{BWEN16} in the case of two-component systems we model a Bragg pulse in the Raman-Nath limit \cite{M09,DB01}, in which the motion of the particles
during the pulse is neglected, with the Bragg pulse operator
\be
U_B(q,A)=e^{-i A\int \, dx \cos (qx) \left(\Psi^\dagger_\uparrow(x)\Psi_\uparrow(x)+\Psi^\dagger_\downarrow(x)\Psi_\downarrow(x)\right)}\, .
\ee
The action of such an instantaneous pulse on an arbitrary eigenstate of the Hamiltonian (\ref{ham}) is given by
\be
|\Phi_{N,M}^{q,A}(\bj,\bl)\rangle=U_B(q,A)|\Phi_{N,M}(\bj,\bl)\rangle\, ,
\ee
with $|\Phi_{N,M}^{q,A}(\bj,\bl)\rangle$ given by (\ref{eigen}) with the wavefunction multiplied by $e^{-i A\sum_{k=1}^N\cos(q x_k)}$.
This means that the effect of the Bragg pulse is that in the Slater determinant describing the charge degrees of freedom we have
to replace $\phi_j(x)$ with $\phi_j(x)e^{- i A\cos(qx)}$. After the pulse the time evolution is given by the Hamiltonian (\ref{ham}) with
$V(x)=m\omega^2 x^2/2$ and the dynamics of the single particle eigenfunctions can be computed analytically using the propagator of the
quantum harmonic oscillator
\be
K(x,u|\, t)=\left(\frac{m \omega}{2\pi i \sin(\omega t)}\right)^{1/2}\exp\left(\frac{-m\omega(x^2+u^2)\cos(\omega t)+2m\omega x u}{2 i
\sin (\omega t)}\right)\, ,
\ee
and $\phi_j(x,t)=\inti K(x,u|\,t)e^{-i A\cos(q x)}\phi_j(u)\,  du\, $. One obtains \cite{BWEN16}
\begin{align}\label{i20}
\phi_j(x,t)&=\sum_{n=-\infty}^\infty I_n(-i A)e^{-i nq \cos(\omega t)\left(x+n q\frac{\sin(\omega t)}{2 m \omega}\right)}
\phi_j\left(x+n q\frac{\sin(\omega t)}{m\omega}\right)e^{-i\omega\left(j+\frac{1}{2}\right) t}\, ,
\end{align}
with $I_n(x)=\int_0^{\pi} e^{x\cos\theta}\cos(n\theta)d\theta/\pi$ the modified Bessel function of the first kind. Therefore, the dynamics of the
GY model in the quantum Newton's cradle is given by the determinant representation (\ref{statictfull}) with matrices (\ref{statict}) at finite
temperature and  (\ref{statict0}) at zero temperature with the time-evolved single particle orbitals defined in (\ref{i20}).
\begin{figure}
\includegraphics[width=1\linewidth]{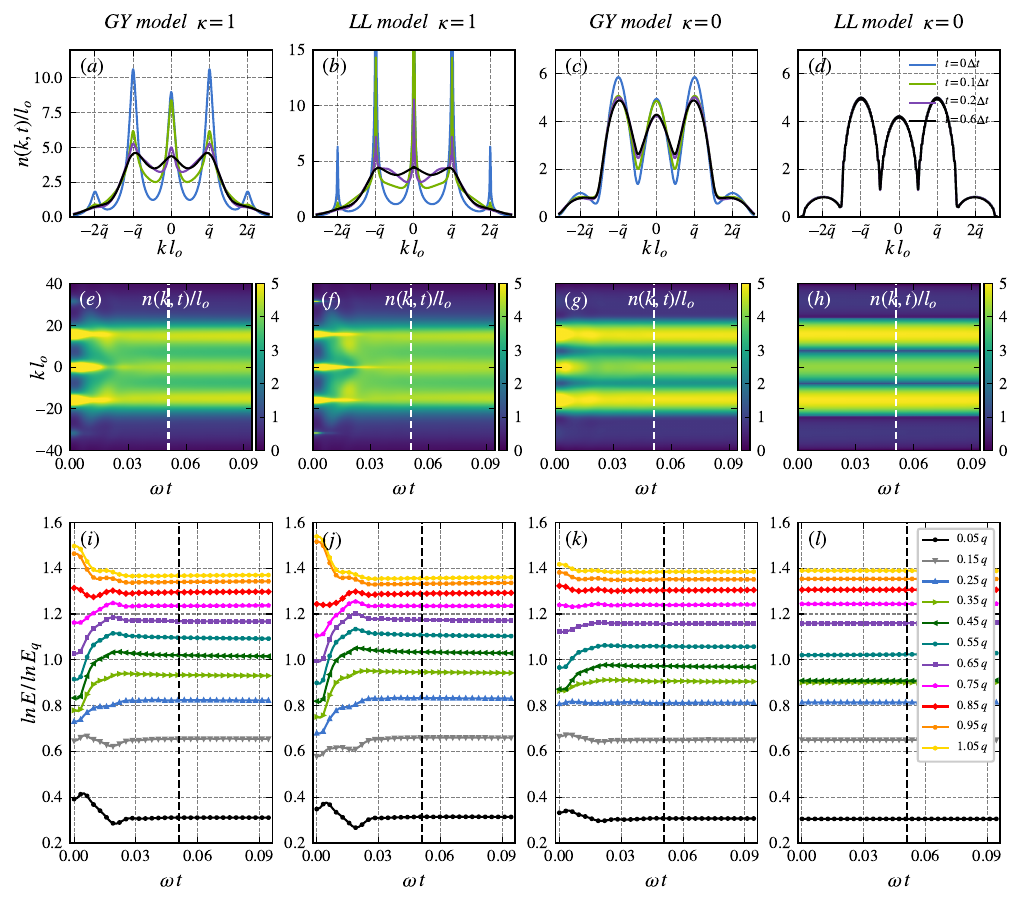}
\caption{First row: The momentum distribution functions after the Bragg pulse at $t=\{0\, ,0.1\Delta t\, ,0.2\Delta t\, ,0.6\Delta t\}$ ($\Delta t=\pi/40\omega$) for systems of $N=32$ particles at zero temperature ($\omega=1$, $A=1.5$, $q=5\pi$, $l_0=1$, $\tilde q= ql_0$). 
We present results for the bosonic GY model a), bosonic LL model b), fermionic GY model c) and single component free fermions (anyonic LL 
model at $\kappa=0$) d). The results for the GY model were computed using  Eqs.~(\ref{statictfull}), (\ref{statict0}) and (\ref{i20}).
Second row: Time evolution of the momentum distribution functions showing the
rapid population of the modes between the $\pm q$ satellites due to hydrodynamization  which is essentially complete by
$T_{hd}=2\pi/\omega_{hd}$ (marked by the white dashed line). Third row: Time evolution of the integrated energy (in units of $E_q=q^2/2m$)
in $0.1 q$ wide momentum groups. The average momentum  of each groups is shown in the legend. The dashed black line marks $T_{hd}$.
}
\label{fighd}
\end{figure}
In the TG regime the time evolution of the GY model in the QNC setup is periodic with period $\pi/\omega$. This statement can proved using
the relation $I_n(-iA)=I_{-n}(iA)$ (see the integral representation) in (\ref{i20}) resulting in  $\phi_j(x,t+\pi/\omega)=e^{-i(j+1/2)\pi}
\phi_j(x,t)$. The correlators involve products of wavefunctions of the type $\overline{\chi}_{N,M} \chi_{N,M}$ resulting in cancellation of
the phases and therefore the densities and momentum distributions are periodic with period $\pi/\omega$. In the single component case Berg
{\it et al.} \cite{BWEN16} showed that there are two separate time scales in the problem: rapid and trap-insensitive dephasing after the
pulse followed by the slow periodic behaviour. The fastest time scale is associated with hydrodynamization and was  experimentally observed
in single component TG bosons \cite{LZGR23}. Below we will investigate the hydrodynamization  in the GY model and highlight the differences
between the single and two-component case.

Hydrodynamization occurs in systems which are quenched with energies much larger than the ground-state energy and is  characterized by a
rapid onset of hydrodynamics before local thermal equilibrium is established \cite{FHS18}. Hydrodynamization  takes place on the fastest
available timescale which is related to the Bragg peak energies  and can be seen in the redistribution of energy
among distant momentum modes \cite{LZGR23}. For Bragg pulses with $A\sim 1$ the hydrodynamization frequency $\omega_{hd}$ can be obtained from
the difference of the $n=0$ and $n=\pm 1$ Bragg orders $\omega_{hd}=q^2/2m$ and the associated timescale of hydrodynamization is given by
$T_{hd}=2\pi/\omega_{hd}$.

In order to investigate the action of the Bragg pulse on the MDF and the subsequent time evolution it is useful to remind the reader some
analytical results on the correlators of TG gases. For any interaction and geometry the following relation is valid [$\Phi\equiv\Phi_{N,M}
(\bj,\bl)$]
\begin{align}
\langle\Phi|U_B^\dagger(q,A)\Psi^\dagger_\sigma(x)\Psi_\sigma(y)U_B(q,A)|\Phi\rangle=
e^{-2 i A\sin\left(q\frac{x-y}{2}\right)\sin\left(q\frac{x+y}{2}\right)}\langle\Phi|\Psi^\dagger_\sigma(x)\Psi_\sigma(y)|\Phi\rangle
\end{align}
which can be proved by using the explicit expression of the mean value on terms of the wavefunctions and the fact that the action of the
Bragg operator multiplies the wavefunction of an arbitrary state with $e^{-i A\sum_{k=1}^N\cos(q x_k)}$. Then, performing similar calculations
like in the Supplemental Material of \cite{BWEN16}, one can show that in the case of circular geometry the momentum distribution function after
the pulse is given by
\be\label{i21}
n_\sigma(k,t=0)=\sum_{l=-\infty}^\infty c_l(A) n_\sigma^{(0)}(k+l q)\, ,
\ee
where $n_\sigma^{(0)}(k)$ is the MDF before the pulse and the coefficients $c_l(A)$ depend on the value of $A$. Eq.~(\ref{i21}) shows that in
the case of a homogeneous system the MDF after the pulse is a sum of copies of the ground-state MDF at $T=0$ (thermal MDF at finite temperature)
centered around multiples of $q$. Using the Local Density Approximation one expects that a similar picture holds in the case of weak harmonic
trapping. Therefore, it is useful to study the MDF of homogeneous systems. In the case of single component anyons without trapping the large
distance asymptotics of the field-field correlators is given by \cite{CM07,CS09,P19}
\be\label{asymsingle}
g^{(-)}(x,0)\equiv\langle\Psi^\dagger(x)\Psi(0)\rangle\sim a\,\frac{e^{i k_F(\kappa-1) x}}{x^{1-\kappa+\frac{\kappa^2}{2}}}
+b\,\frac{e^{i k_F(\kappa+1) x}}{x^{1+\kappa+\frac{\kappa^2}{2}}}\, ,\ \ x>0\, ,
\ee
with $a,b$ constants that can be found in \cite{P19}. Note that $g^{(-)}(-x,0)=\overline{g^{(-)}(x,0)}$. We focus on the large distance asymptotics
because via Fourier transform they give the behaviour of the MDF for $k\sim 0$. In the bosonic case, $\kappa=1$, the first term in the right hand-side
of (\ref{asymsingle}) is dominant and we have $g^{(-)}(x,0)\sim a/x^{1/2}$ which results in an MDF behaving like $n(k)\sim 1/k^{1/2}$ for
$k\rightarrow 0$ \cite{VT79a,VT79b,JMMS80,G04}. For free fermions, $\kappa=0$, both terms are relevant and they reproduce the well known result
$g^{(-)}(x,0)=\sin(k_F x)/\pi x$  with $n(k)\sim \bm{1}_{[-k_F,k_F]}$.

The large distance asymptotics for homogeneous impenetrable Gaudin-Yang anyons is given by \cite{P19} ($\nu=-i \frac{\ln 2}{2\pi}-\frac{\kappa}{2}$)
\be\label{asymdouble}
g_\sigma^{(-)}(x,0)\sim\frac{e^{-2 i\nu k_F x}}{x^{2\nu^2+1}} \left[a\, x^{-2\nu} e^{- i k_F x}+ b\, x^{2\nu} e^{ i k_F x}\right]\, ,\ \ x>0\, ,
\ee
\begin{figure}
\includegraphics[width=1\linewidth]{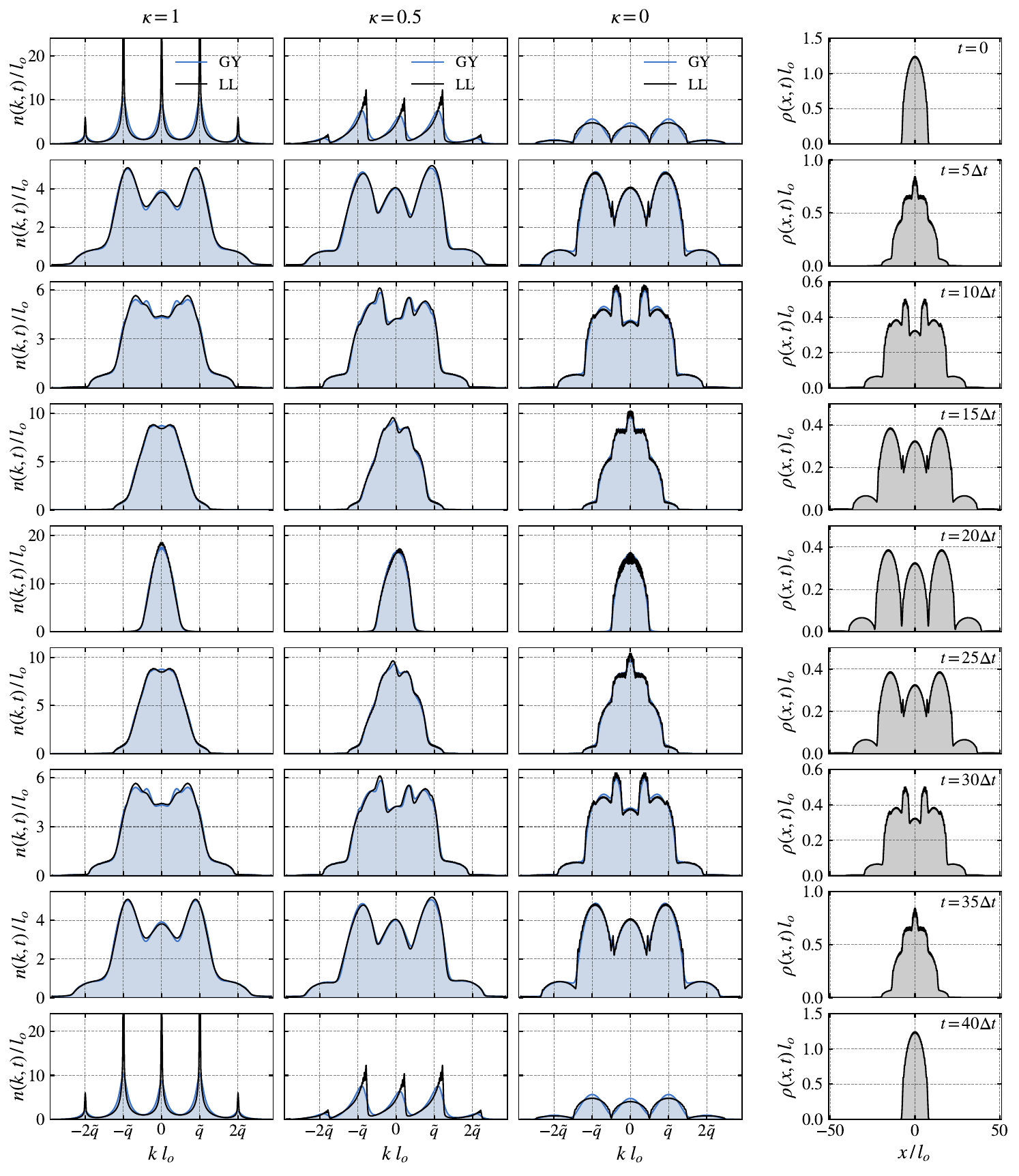}
\caption{Momentum distribution functions and densities of the GY and LL anyonic models for $t=\{0,5,10,15,20,25,30,35,40\}\times \Delta t\, ,$ $\Delta t=
\pi/40 \omega$ in the QNC setup a for a system of $N=30$  particles at zero temperature  ($\omega=1$, $A=1.5$, $q=5\pi$, $l_0=1$, $\tilde q= ql_0$).
First column $\kappa=1$ (bosons), second column $\kappa=0.5$, third column $\kappa=0$ (fermions). Fourth column: Density which is the same for all
values of $\kappa$ and for both GY and LL models. The quantities for the GY model are computed using Eqs.~(\ref{statictfull}), (\ref{statict0}) and (\ref{i20}).
}
\label{figqnc}
\end{figure}
with $a,b$ constants which can be found in \cite{P19}. Similar to the single component case in the bosonic case the first term in the right hand-side
is dominant obtaining \cite{CSZ05}
\be\label{i22}
g_\sigma^{(-)}(x,0)\sim e^{-\frac{\ln 2}{\pi} k_F x}x^{-\frac{1}{2}+\frac{1}{2}\left(\frac{ln 2}{\pi}\right)^2}\, ,
\ee
and in the fermionic case both terms contribute with the results \cite{BL87,B91,CZ04a,CZ04b,FB04}
\be\label{i23}
g_\sigma^{(-)}(x,0)\sim e^{-\frac{\ln 2}{\pi} k_F x}x^{-1+\frac{1}{2}\left(\frac{ln 2}{\pi}\right)^2}\sin\left(k_F x-\ln 2 \ln x/\pi-\varphi_0\right)\, ,
\ee
with $\varphi_0$ a constant. The main feature of the asymptotics (\ref{i22}) and (\ref{i23}) is the presence of the exponential decreasing term
$e^{-\frac{\ln 2}{\pi} k_F x}$ even though we are at zero temperature. This is  a general feature of multicomponent systems in the
spin-incoherent regime:  in the case of a system with $M$  components the exponential terms is $e^{-\frac{\ln M}{\pi} k_F x}$ \cite{FB04}. The algebraic
corrections are very close to the ones for single component systems $\frac{1}{2}\left(\frac{ln 2}{\pi}\right)^2\sim 0.024$ but in the fermionic case
the oscillatory term has a $\ln x$ term dependence in addition to a phase.  In the bosonic case this results in a MDF for the GY model which is wider and
does not present the weak singularity $k^{-1/2}$ characteristic of single component bosons. In the fermionic case the opposite statement is true with the
MDF for the GY model being narrower than the similar quantity for free fermions. These observations remain valid also in the case of harmonic trapping
as it can be seen in the first row of Fig.~\ref{fighd} where we present the MDF for several values of $t$ immediately after the Bragg pulse.
One can see that the MDF for free fermions (their momenta are just the rapidities) remains almost unchanged while in the other cases we can clearly
see the transfer of energy from the $\pm q, \pm 2q$ satellites to the modes between the  peaks. The time evolution of the MDF is shown in the second
row of Fig.~\ref{fighd} where it can be seen clearly that the modes between the first Bragg peaks are populated very rapidly. This is due to the
fact that these modes  are composed of the widest range of rapidities and, hence, they dephase fastest. One can see that the process of hydrodynamization
takes place on the timescale set by $T_{hd}=2\pi/\omega_{hd}$. The rapid change in the energy distribution associated with hydrodynamization can be seen
more clearly by integrating the kinetic energy in successive momentum ranges and plotted as function of time as it can be seen in the last row of
Fig.~\ref{fighd}. Each curve presents the time evolution of the integrated energy in $0.1 q$ wide momentum groups up to the first Bragg peak. For the
GY model one can see that the rapid initial change in the intermediate momentum groups is more dramatic in the bosonic case compared with the fermionic
case which is to be expected due to the wider initial MDF in the fermionic case. The change is also more pronounced in the bosonic LL model compared with
the bosonic GY model due to the quasi-condensate nature of the MDF in the single component case compared with the wider MDF of the spin-incoherent GY model.
While the changes in the free fermionic case are extremely small it should be noted that they are nonzero.

In Fig.~\ref{figqnc}  we present results the time evolution of the MDF and densities for a system of $N=30$ particles at zero temperature
for both the GY model and its single component counterpart and three values of the statistics parameter: $\kappa=1$ (bosons), $\kappa=0.5$ and
$\kappa=0$ (fermions). For $\kappa=0.5$  one can see the nonsymmetric  momentum  distribution and that at $t=0$ the MDF of the single component
is narrower (the leading term comes from the Fourier transform of $e^{i k_F(\kappa-1) x}/x^{1-\kappa+\frac{\kappa^2}{2}}$) than the one for the
two-component system. The nonsymmetry remains  visible for the entire period of the oscillation. From the first three columns of Fig.~\ref{figqnc}
we see that the overlap between the MDF for single and two-component systems during the oscillations is pretty large with significant differences
occurring in the vicinities of $t=p \pi/\omega$ and $t=\frac{1}{2} p \pi/\omega$ with $p$ integer. However, we should point out that the tails of
the MDFs which behave like $n(k,t)\sim C(t)/k^4$ with $C(t)$ the Tan contact are different with the contacts for the two-component systems
being smaller than the contacts for the single component ones \cite{CSZ05}. The density for systems with the same number of particles is independent of
statistics and the number of components in the system. Its dynamics is shown in the fourth column of Fig.~\ref{figqnc} where it can be seen that
it oscillates in out-of-phase with respect to the MDFs: the density is narrowest when the MDF is largest and the converse is also true.

\section{Conclusions}\label{s8}

In this paper we have investigated the nonequilibrium dynamics of the Gaudin-Yang model in two experimentally accessible scenarios: the quench induced by
the sudden change in the trap's frequency and the quantum Newton's cradle setup. Our investigation used a determinant representation for the
space-, time-, and temperature-dependent correlators which is extremely easy to implement numerically with the main computational effort coming from the
calculation of the partial overlaps for the time evolved single particle orbitals. When the model is subjected to a quench of the trap's frequency  we have
identified a collective many-body bounce effect with an amplitude that depends on the statistics of the particles and for the QNC setup we have performed
a thorough study of the dynamics and hydrodynamization. A natural extension of our work would be the derivation of similar representations for the lattice
analog of the GY model, the Hubbard model. This will be deferred to a future publication.

\acknowledgments
Financial support  from the Grant No. 30N/2023 of the National Core Program  of the Romanian Ministry of Research, Innovation and Digitization is gratefully
acknowledged.

\appendix

\section{Derivation of the determinant representation for the form factors}\label{aff}

In this Appendix we will derive the determinant representations for the form factors (\ref{formff}). The arbitrary state in the $(N+1,M)$-sector
appearing in the definition of the form factor is characterized by $\bj=(j_1,\cdots,j_{N+1})$ which describes the charge degrees of freedom and
$\bl=(\lambda_1,\cdots,\lambda_M)$ specifying the spin sector with $e^{i\lambda_a (N+1)}=(-1)^{M-1}$ for $a=1,2,\cdots,M$. The other state appearing
in the definition of the form factor belongs to the  $(N,\bar{M})$-sector and is characterized by $\bq=(q_1,\cdots,q_N)$ and $\bmu=(\mu_1,\cdots,
n_{\bar{M}})$ with $e^{i\mu_b N}=(-1)^{\bar{M}-1}$ for $b=1,2,\cdots,\bar{M}$. Introducing
\be
\Lambda=\sum_{a=1}^M \lambda_a\, ,\ \Theta=\sum_{b=1}^{\bar{M}}\mu_b\,  \mbox{ and } \omega=e^{i\Lambda}\, ,\ \nu=e^{i \Theta}\, ,
\ee
we have
\begin{align}\label{cyclic}
\eta_{N+1,M}^{(\ba,\alpha_1\alpha_2 \cdots \alpha_{N+1})}(\bl)&=\omega\, \eta_{N+1,M}^{(\ba,\alpha_2\alpha_3\cdots\alpha_{N+1}\alpha_1)}(\bl)\, ,\\
\eta_{N,\bar{M}}^{(\ba,\alpha_1\alpha_2 \cdots \alpha_{N})}(\bmu)&=\nu\, \eta_{N,\bar{M}}^{(\ba,\alpha_2\alpha_3\cdots\alpha_{N}\alpha_1)}(\bmu)\, ,
\end{align}
which is a consequence of the fact that the $XX$ spin chain wavefunctions (\ref{wavefs}) are also eigenfunctions of the cyclic shift operator on the
lattice. Also, using (\ref{i1}) we obtain
\be\label{i3}
\mathcal{F}_{N,M}^{(\sigma)}(\bj,\bl;\bq,\bmu|x,t)=e^{i t\mu_\sigma} \langle\Psi_{N,\bar{M}}(t|\bq,\bmu)|\Psi_\sigma(x)|\Phi_{N+1,M}(t|\bj,\bl)
\rangle\, ,
\ee
with the time evolved wavefunctions given by (\ref{waveft}) and $\mu_\uparrow=\mu-B\, , \mu_\downarrow=\mu+B$. Using the commutation relations
(\ref{comm}) and the symmetry of the wavefunctions (\ref{gensym}) the starting point of our calculations is
\begin{align}
\mathcal{F}_{N,M}^{(\sigma)}(x,t)&=(N+1)!e^{i t\mu_\sigma} \int_{L-}^{L+}\prod_{i=1}^N dx_i\, \sum_{\alpha_1,\cdots,\alpha_N=\{\uparrow,\downarrow\}
}^{[N,\bar{M}]}
\overline{\chi}_{N,\bar{M}}^{\ba}(x_1,\cdots,x_N,t|\bq,\bmu)\chi_{N+1,M}^{\ba\sigma}(x_1,\cdots,x_N,x,t|\bj,\bl)
\end{align}
where the bar denotes complex conjugation and $L_\pm$ are the limits of integration which can differ depending on the system we consider. For example, in the
case of trapping we have $L_\pm=\pm\infty$ but in the case of Dirichlet boundary conditions in the box $[0,L]$ we have $L_-=0$ and $L_+=L$. The evolved
wavefunctions are
\begin{align}
\overline{\chi}_{N,\bar{M}}^{\ba}(x_1,\cdots,x_N,t|\bq,\bmu)&=\frac{1}{N!N^{\bar{M}/2}}\left[\sum_{R\in S_N}\theta(R\boldsymbol{x})e^{-i\frac{\pi\kappa}{2}\sum_{1\le a<b<N}\mbox{\small{sign}}(x_a-x_b)}
\overline{\eta}_{N,\bar{M}}^{(\ba,R\ba)}(\bmu)\right]\nonumber\\
&\qquad\qquad\qquad\times \sum_{Q\in S_N}(-1)^Q\prod_{l=1}^N\overline{\phi}_{q_{Q(l)}}(x_l,t)\, ,
\end{align}
and
\begin{align}
\chi_{N+1,M}^{\ba\sigma}(x_1,\cdots,x_N,x,t|\bj,\bl)&=\frac{1}{(N+1)!(N+1)^{M/2}}\left[\sum_{R'\in S_{N+1}}\theta(R'\boldsymbol{x}' )e^{i\frac{\pi\kappa}{2}\sum_{1\le a<b<N}\mbox{\small{sign}}(x_a-x_b)}
e^{i\frac{\pi\kappa}{2}\sum_{a=1}^N\mbox{\small{sign}}(x_a-x)}\right.\nonumber\\
&\qquad\qquad\qquad\times \left.\eta_{N+1,M}^{(\ba\sigma,R'\ba\sigma)}(\bl)\right]
 \sum_{P\in S_{N+1}}(-1)^P\prod_{l=1}^N\phi_{j_{P(l)}}(x_l,t)\phi_{j_{P(N+1)}}(x,t)\, ,
\end{align}
where $R\boldsymbol{x}=x_{R(1)}<\cdots<x_{R(N)}$ and $R'\boldsymbol{x}'=x_{R'(1)}<\cdots<x_{R'(N+1)}$ (one $x_{R'(i)}=x$).
Multiplying the wavefunctions we encounter products of the type $\theta(R\boldsymbol{x})\theta(R'\boldsymbol{x}')=
\prod_{j=1}^N\delta_{R(j),R'(j)}\theta(R'\boldsymbol{x}')$ with $R\in S_N$ and  $R'\in S_{N+1}$.
The $(N+1)!$ surviving terms can be divided in $N+1$ sets of $N!$ terms depending on the position of $R'(N+1)$ which indexes the position
of $x$ i.e.,
\begin{align}
\sum_{R\in S_N}\sum_{R'\in S_{N+1}}\theta(R\boldsymbol{x})\theta(R'\boldsymbol{x}')&=
\sum_{R\in S_N}
\left\{\theta(x_{R(1)}<\cdots<x_{R(N)}<x))\right.\nonumber\\
&\qquad\qquad+\sum_{n=1}^{N-1}\theta(x_{R(1)}<\cdots<x_{R(n)}<x<x_{R(n+1)}<\cdots<x_{R(N)})\nonumber\\
&\left.\qquad\qquad+\,\theta(x<z_{R(1)}<\cdots<x_{R(N)})\right\}\, .
\end{align}
If we consider the set in which $x$ is on the $n$-th position, then, for a given  $\boldsymbol{\alpha}=(\alpha_1\cdots\alpha_N)$, the product of
spin wavefunctions is given  by $\overline{\eta}_{N,\bar{M}}^{(\ba,R\ba)}(\bmu) \, \eta_{N+1,M}^{(\ba\sigma,R'\ba\sigma)}(\bl)=
\overline{\eta}_{N,\bar{M}}^{(\ba,\alpha_1\cdots\alpha_N)}(\bmu)\, \eta_{N+1,M}^{(\ba\sigma,\alpha_1\cdots\alpha_n\sigma\alpha_{n+1}\cdots\alpha_N)}(\bl)$.
Collecting these results (\ref{i3}) can be written as
\begin{align}
\mathcal{F}_{N,M}^{(\sigma)}(x,t)&=\frac{e^{i t\mu_\sigma}}{N!N^{\bar{M}/2}(N+1)^{M/2}}\int_{L_-}^{L_+}\prod_{i=1}^N dx_i\sum_{R\in S_N}
\left\{\theta(x_{R(1)}<\cdots<x_{R(N)}<x)e^{-i\frac{\pi\kappa N}{2}}F_\sigma(N)\right.\nonumber\\
&\qquad\qquad+\sum_{n=1}^{N-1}\theta(x_{R(1)}<\cdots<x_{R(n)}<x<x_{R(n+1)}<\cdots<x_{R(N)})\,e^{-i \frac{\pi\kappa}{2} n}e^{i\frac{\pi\kappa}{2}(N-n)}F_\sigma(n)\nonumber\\
&\left.\qquad\qquad+\,\theta(x<z_{R(1)}<\cdots<x_{R(N)})\,e^{ i\frac{\pi\kappa}{2} N} F_\sigma(0)\right\}
\left(\sum_{Q\in S_N}(-1)^Q\prod_{l=1}^N\overline{\phi}_{q_{Q(l)}}(x_l,t)\right)\nonumber\\
&\qquad\qquad\qquad\times \left( \sum_{P\in S_{N+1}}(-1)^P\prod_{l=1}^N\phi_{j_{P(l)}}(x_l,t)\phi_{j_{P(N+1)}}(x,t)\right)\, ,
\end{align}
with
\be
F_\sigma(n)=\sum_{\alpha_1,\cdots,\alpha_N=\{\uparrow,\downarrow\}}^{[N,\bar{M}]}\overline{\eta}_{N,\bar{M}}^{(\ba,\alpha_1\cdots\alpha_N)}(\bmu)
\, \eta_{N+1,M}^{(\ba\sigma,\alpha_1\cdots\alpha_n\sigma\alpha_{n+1}\cdots\alpha_N)}(\bl)\, .
\ee
Using the cyclic property of the spin wavefunctions (\ref{cyclic}) and the fact that $\overline{\omega}=\omega^{-1}$, $\overline{\nu}=\nu^{-1}$ we find that
\be
F_\sigma(n)=(\overline{\omega}\nu)^{N-n}F_\sigma(N)\, ,\ \ F_\sigma\equiv F_\sigma(N)\, ,
\ee
and, therefore,
\begin{align}\label{i4}
\mathcal{F}_{N,M}^{(\sigma)}(x,t)&=\frac{e^{i t\mu_\sigma}e^{-i\frac{\pi\kappa N}{2}}F_\sigma}{N!N^{\bar{M}/2}(N+1)^{M/2}}\int_{L_-}^{L_+}\prod_{i=1}^N dx_i\sum_{R\in S_N}
\left\{\theta(x_{R(1)}<\cdots<x_{R(N)}<x)\right.\nonumber\\
&\qquad\qquad+\sum_{n=1}^{N-1}\theta(x_{R(1)}<\cdots<x_{R(n)}<x<x_{R(n+1)}<\cdots<x_{R(N)})\,\left(\overline{\omega}\nu e^{i \pi\kappa} \right)^{N-n}\nonumber\\
&\left.\qquad\qquad+\,\theta(x<z_{R(1)}<\cdots<x_{R(N)})\,\left(\overline{\omega}\nu e^{i \pi\kappa} \right)^{N}\right\}
\left(\sum_{Q\in S_N}(-1)^Q\prod_{l=1}^N\overline{\phi}_{q_{Q(l)}}(x_l,t)\right)\nonumber\\
&\qquad\qquad\qquad\times \left( \sum_{P\in S_{N+1}}(-1)^P\prod_{l=1}^N\phi_{j_{P(l)}}(x_l,t)\phi_{j_{P(N+1)}}(x,t)\right)\, .
\end{align}
Fortunately, one can show that
\begin{align}\label{identity1}
\sum_{R\in S_N} &\left\{ \theta(x_{R(1)}<\cdots<x_{R(N)}<x) +\sum_{n=1}^{N-1}\theta(x_{R(1)}<\cdots<x_{R(n)}<x<x_{R(n+1)}<\cdots<x_{R(N)}) \left(\overline{\omega}\nu e^{i \pi\kappa} \right)^{N-n}\right. \nonumber\\
&\left.\qquad\qquad+ \theta(x<z_{R(1)}<\cdots<x_{R(N)})\,\left(\overline{\omega}\nu e^{i \pi\kappa} \right)^{N}\right\}
=\prod_{n=1}^N\rho(x-x_n)\, ,
\end{align}
with
\be
\rho(x)=\theta(x)+e^{i\pi\kappa}\overline{\omega}\nu \theta(-x)\, .
\ee
This identity is valid for all $x_1,\cdots,x_N\in[L_-,L_+]$ when the $x_i$'s are different. The value of $\rho(0)$ is not important because when
two coordinates are equal the determinants in the right hand-side of (\ref{i4}) vanish. In order to prove this identity it is instructive to look
at the particular case $N=2$ where the left hand side of (\ref{identity1}) is
\begin{align}\label{aj1}
L.H.S=\sum_{R\in S_2}\left \{ \theta(x_{R(1)}<x_{R(2)}<x)+\theta(x_{R(1)}<x<x_{R(2)})\, \overline{\omega}\nu e^{i \pi\kappa}+\theta(x<x_{R(1)}<x_{R(2)})\left(\,\overline{\omega}\nu e^{i \pi\kappa} \right)^2   \right\}
\end{align}
and the right hand side is
\begin{align}\label{aj2}
R.H.S.=\theta(x-x_1)\theta(x-x_2)+\left[\theta(x-x_1)\theta(x_2-x)+\theta(x-x_2)\theta(x_1-x)\right] \, \overline{\omega}\nu e^{i \pi\kappa}
+\theta(x_1-x)\theta(x_2-x) \left(\,\overline{\omega}\nu e^{i \pi\kappa} \right)^2\, .
\end{align}
The equality of (\ref{aj1}) and (\ref{aj2}) becomes evident noticing that $\theta(x-x_1)\theta(x-x_2)=\sum_{R\in S_2}\theta(x_{R(1)}<x_{R(2)}<x)$,
$\left[\theta(x-x_1)\theta(x_2-x)+\theta(x-x_2)\theta(x_1-x)\right]= \sum_{R\in S_2} \theta(x_{R(1)}<x<x_{R(2)})$ and
$\theta(x_1-x)\theta(x_2-x)=\sum_{R\in S_2}\theta(x<x_{R(1)}<x_{R(2)})$. The general case is proved along the same lines by noticing that the terms
multiplied by $\left(\,\overline{\omega}\nu e^{i \pi\kappa} \right)^{N-n} $ obtained by expanding the r.h.s of (\ref{identity1})  are equal to
$ \sum_{R\in S_N} \theta(x_{R(1)}<\cdots<x_{R(n)}<x<x_{R(n+1)}<\cdots<x_{R(N)})$.

Inserting the identity (\ref{identity1}) in (\ref{i4}) we see that the integration over the charge degrees of freedom
can be written in a factorized form
\begin{align}
\mathcal{F}_{N,M}^{(\sigma)}(x,t)&=\frac{e^{i t\mu_\sigma}e^{-i\frac{\pi\kappa N}{2}}F_\sigma}{N!N^{\bar{M}/2}(N+1)^{M/2}}\int_{L_-}^{L_+}\prod_{i=1}^N dx_i \rho(x-x_i)
\sum_{Q\in S_N}\sum_{P\in S_{N+1}}(-1)^{P+Q}\left(\prod_{l=1}^N\overline{\phi}_{q_{Q(l)}}(x_l,t) \phi_{j_{P(l)}}(x_l,t)\right)
\nonumber\\
&\qquad\qquad\qquad\qquad\qquad\qquad\times\phi_{j_{P(N+1)}}(x,t)\, .
\end{align}
Using the orthonormality of the wavefunctions $\int_{L_-}^{L_+}\overline{\phi}_q(v,t)\phi_j(v,t)\, dv=\delta_{j,q}$  the integrals over $x_i$ can be calculated
using the formula
\begin{align}
\int_{L_-}^{L_+}\rho(x-v)\overline{\phi}_q(v,t)\phi_j(v,t)\, dv&=\int_{L_-}^x \overline{\phi}_q(v,t)\phi_j(v,t)\, dv+
e^{i\pi\kappa} \overline{\omega}\nu \int_x^{L_+}\overline{\phi}_q(v,t)\phi_j(v,t)\, dv\, ,\nonumber\\
&= \int_{L_-}^{L_+} \overline{\phi}_q(v,t)\phi_j(v,t)\, dv -\left(1-  e^{i\pi\kappa} \overline{\omega}\nu\right) \int_x^{L_+}\overline{\phi}_q(v,t)\phi_j(v,t)\, dv\, ,\nonumber\\
&=f(j,q|x,t)\, ,\nonumber
\end{align}
with
\be\label{deff}
f(j,q|x,t)=\delta_{j,q}-\left(1-e^{i\pi\kappa} \overline{\omega}\nu\right)\int_x^{L_+}\overline{\phi}_q(v,t)\phi_j(v,t)\, dv\, .
\ee
Therefore, we find
\begin{align}
\mathcal{F}_{N,M}^{(\sigma)}(x,t)&=\frac{e^{i t\mu_\sigma}e^{-i\frac{\pi\kappa N}{2}}F_\sigma}{N!N^{\bar{M}/2}(N+1)^{M/2}}
\sum_{Q\in S_N}\sum_{P\in S_{N+1}}(-1)^{P+Q}\left(\prod_{l=1}^N f(j_{P(l)},q_{Q(l)}|x,t) \right)\phi_{j_{P(N+1)}}(x,t)\, ,
\end{align}
with the last part which can be written as a determinant
\begin{align}
\sum_{Q\in S_N} (-1)^Q\left|
\begin{array}{cccc}
f(j_1,q_{Q(1)}) &\cdots &f(j_1,q_{Q(N)}) & \phi_{j_1}\\
\vdots &\ddots &\vdots & \vdots\\
f(j_{N+1},q_{Q(1)})&\cdots&f(j_{N+1},q_{Q(N)}) &  \phi_{j_{N+1}}
\end{array}
\right|(x,t)\, .
\end{align}
Reorganizing the columns such that $(Q_1,\cdots,Q_N)\rightarrow (1,\cdots,N)$ gives a $(-1)^Q$ sign, therefore, the sum produces $N!$ identical
terms and the form factor can be written as
\begin{align}
\mathcal{F}_{N,M}^{(\sigma)}(x,t)&=\frac{e^{i t\mu_\sigma}e^{-i\frac{\pi\kappa N}{2}}}{N^{\bar{M}/2}(N+1)^{M/2}}F_\sigma \det_{N+1} D(\bj,\bq|x,t)\, ,
\end{align}
with $D(\bj,\bq|x,t)$ a square matrix of dimension $N+1$ and elements
\begin{align}
[D(\bj,\bq|x,t)]_{ab}=\left\{
\begin{array}{lll}
f(j_a,q_b|x,t) & \mbox{ for } & a=1,\cdots,N+1\, ; \, b=1,\cdots,N\, ,\\
\phi_{j_a}(x,t) & \mbox { for } & a=1,\cdots,N+1\, ;\,  b=N+1\, .
\end{array}
\right.
\end{align}

The only thing that remains is to compute is the $F_\sigma$ factor. We start with the case $\sigma=\uparrow$. Taking into account that
the $n$'s are the position of the spin down particles on the lattice the first observation that we make is that the sum over $\alpha$'s in
\be
F_\sigma\equiv F_\sigma(N)=\sum_{\alpha_1,\cdots,\alpha_N=\{\uparrow,\downarrow\}}^{[N,\bar{M}]}\overline{\eta}_{N,\bar{M}}^{(\ba,\alpha_1\cdots\alpha_N)}(\bmu)
\, \eta_{N+1,M}^{(\ba\sigma,\alpha_1\cdots\alpha_N\sigma)}(\bl)\, .
\ee
is equivalent with $\sum_{1\le n_1<\cdots<n_N\le N}$. For $\sigma=\uparrow$ the product $\overline{\eta}_{N,\bar{M}}\eta_{N+1,M}$ is symmetric in $n$'s
(the products of sign factors cancel) and vanish when two of them are equal. Therefore, we have
\be
\sum_{1\le n_1<\cdots<n_M\le N}=\frac{1}{M!}\sum_{n_1=1}^N\cdots\sum_{n_M=1}^N\, ,
\ee
and
\begin{align}
F_\uparrow&=\frac{1}{M!}\sum_{n_1=1}^N\cdots\sum_{n_M=1}^N\left(\sum_{Q\in S_M}(-1)^Q\prod_{k=1}^M e^{- i n_k \mu_{Q(k)}}\right)
\left(\sum_{P\in S_M}(-1)^P\prod_{k=1}^M e^{ i n_k \lambda_{P(k)}}\right)\, ,\nonumber\\
&=\frac{1}{M!}\sum_{n_1=1}^N\cdots\sum_{n_M=1}^N\sum_{Q\in S_M}\sum_{P\in S_M}(-1)^{P+Q}\prod_{k=1}^M e^{i n_k(\lambda_{P(k)}-\mu_{Q(k)})}\, ,\nonumber\\
&=\sum_{P\in S_M} (-1)^P\prod_{k=1}^M\left(\sum_{n=1}^N e^{i n(\lambda_{P(k)}-\mu_j)}\right)\, ,
\end{align}
which shows that $F_\uparrow=\det_M B_\uparrow(\bl,\bmu)$ where the matrix $B_\uparrow$ has elements.
In the $\sigma=\downarrow$ case we have $\bar{M}=M-1$ and $n_M=N+1$ for the $n$'s in $\eta_{N+1,M}^{(\ba\sigma,\ba\sigma)}(\bl)$. The the product $\overline{\eta}_{N,\bar{M}}\eta_{N+1,M}$
is now symmetric in $M-1$ variables and vanish when two of them are equal. We find
\begin{align}
F_\downarrow&=\frac{1}{(M-1)!}\sum_{n_1=1}^N\cdots\sum_{n_{M-1}=1}^N\left(\sum_{Q\in S_{M-1}}(-1)^Q\prod_{k=1}^{M-1} e^{- i n_k \mu_{Q(k)}}\right)
\left(\sum_{P\in S_{M}}(-1)^P\prod_{k=1}^{M-1} e^{ i n_k \lambda_{P(k)}}\right)e^{i(N+1)\lambda_{P(M)}}\, ,\nonumber\\
&=\frac{1}{(M-1)!}\sum_{n_1=1}^N\cdots\sum_{n_{M-1}=1}^N\sum_{Q\in S_{M-1}}\sum_{P\in S_M}(-1)^{P+Q}\prod_{k=1}^M e^{i n_k(\lambda_{P(k)}-\mu_{Q(k)})}(-1)^{M-1}\, ,\nonumber\\
&=\frac{1}{(M-1)!}\sum_{Q\in S_{M-1}}\sum_{P\in S_M}(-1)^{P+Q}\prod_{k=1}^M\left( \sum_{n=1}^N e^{i n (\lambda_{P(k)}-\mu_{Q(k)})}\right)(-1)^{M-1}\, ,
\end{align}
where in the second line we have used the BAEs $e^{i\lambda_a (N+1)}=(-1)^{M-1}$. The analysis of the last expression is similar with the one for the charge degrees of freedom.
We obtain $F_\downarrow=(-1)^{M-1}\det_M B_\downarrow(\bl,\bmu)$  with matrix  elements

\begin{align}
[B_\downarrow(\bl,\bmu)]_{ab}=\left\{
\begin{array}{lll}
\sum_{n=1}^N e^{i n(\lambda_a-\mu_b)} & \mbox{ for } & a=1,\cdots,M\, ; \, b=1,\cdots,M-1\, ,\\
1  & \mbox { for } & a=1,\cdots,M\, ;\,  b=M\, .
\end{array}
\right.
\end{align}
\section{Derivation of the determinant representation for the correlators}\label{ad}

Here, we present the derivation of the determinant representation for the correlators starting with the  summation of form factors for
the mean values appearing on the right hand side of (\ref{defgm}) and (\ref{defgp}).

\subsection{Determinant representation for $\langle \Phi_{N+1,M}(\bj,\bl)|\Psi_\uparrow^\dagger(x,t)\Psi_\uparrow(y,t')|\Phi_{N+1,M}(\bj,\bl)\rangle$}

Using the determinant formulas for the form factors (\ref{formff}) one can obtain similar representations for the mean value of bilocal operators
appearing in the definition of the correlators (\ref{defgm}) and (\ref{defgp}). In this section we consider $A\equiv\langle
\Phi_{N+1,M}(\bj,\bl)|\Psi_\uparrow^\dagger(x,t)\Psi_\uparrow(y,t')|\Phi_{N+1,M}(\bj,\bl)\rangle$. In this case $\bar{M}=M$ and we have
\begin{align}\label{i5}
A&
=\sum_{\substack{q_1<\cdots<q_N\\  \mu_1<\cdots<\mu_{M}}}
\overline{\mathcal{F}}_{N,M}^{(\uparrow)}(\bj,\bl;\bq,\bmu|x,t)
\mathcal{F}_{N,M}^{(\uparrow)}(\bj,\bl;\bq,\bmu|y,t')\, ,\nonumber\\
&=\sum_{\substack{q_1<\cdots<q_N\\  \mu_1<\cdots<\mu_{M}}}\frac{e^{-i (t-t')\mu_\uparrow}}{(N+1)^M N^M}|\det_{M}B_\uparrow(\bl,\bmu)|^2
\overline{\det_{N+1} D(\bj,\bq|x,t)}\det_{N+1} D(\bj,\bq|y,t')\, .
\end{align}
In (\ref{i5}) the summation over $q$'s is independent on the summation on $\mu$'s and the summands  are symmetric functions independently
in $q$'s and $\mu$'s and vanish when two of them are equal (exchanging two $q$'s is equivalent with transposing two columns in the matrix $D$
and we have a product of $\overline{D} D$, the same argument applies in the case of exchange of two $\mu$'s). Therefore, the summations can be
written as
\be
\sum_{\substack{q_1<\cdots<q_N\\  \mu_1<\cdots<\mu_{M}}}=\frac{1}{N!}\sum_{q_1=1}^\infty\cdots\sum_{q_N=1}^\infty\frac{1}{M!}\sum_{\mu_1}
\cdots\sum_{\mu_M}
\ee
where $\sum_\mu h(\mu)=\sum_{l=1}^N h(\mu_l)$ with $\mu_l=\frac{2\pi}{N}\left(-\frac{N}{2}-\frac{1+(-1)^{N-M}}{4}+l\right)$ for an arbitrary
function $h$.

\subsubsection{Summation over $q_1,\cdots,q_N$}

We focus now on the summation over the $q$'s in (\ref{i5}). We find
\begin{align}\label{i6}
A_q&=\frac{1}{N!}\sum_{q_1=1}^\infty\cdots\sum_{q_N=1}^\infty\overline{\det_{N+1} D(\bj,\bq|x,t)}\det_{N+1} D(\bj,\bq|y,t')\, \nonumber\\
&=\frac{1}{N!}\sum_{q_1=1}^\infty\cdots\sum_{q_N=1}^\infty\sum_{P\in S_{N+1}}\sum_{Q\in S_{N+1}}(-1)^{P+Q}
\left(\prod_{l=1}^N\overline{f}(j_{P(l)},q_l|x,t) f(j_{Q(l)},q_l|y,t')\right)\overline{\phi}_{j_{P(N+1)}}(x,t)\phi_{j_{Q(N+1)}}(y,t')\, ,\nonumber\\
&=\frac{1}{N!}\sum_{q_1=1}^\infty\cdots\sum_{q_N=1}^\infty\sum_{R,Q\in S_{N+1}}(-1)^{R}
\left(\prod_{l=1}^N\overline{f}(j_{RQ(l)},q_l|x,t) f(j_{Q(l)},q_l|y,t')\right)\overline{\phi}_{j_{RQ(N+1)}}(x,t)\phi_{j_{Q(N+1)}}(y,t')\, ,
\end{align}
where in the last line we have used the fact that every permutation $P$  can be written as $P=RQ$ with $R$ another permutation. The sum over
permutations in (\ref{i6}) can be written as a sum over determinants
\begin{align*}
\sum_{Q\in S_{N+1}}\left|
\begin{array}{cccc}
\overline{f}(j_{Q(1)},q_1|x,t) f(j_{Q(1)},q_1|y,t') &\cdots & \overline{f}(j_{Q(1)},q_N|x,t) f(j_{Q(N)},q_N|y,t') &
\overline{\phi}_{j_{Q(1)}}(x,t)\phi_{j_{Q(N+1)}}(y,t')\\
\vdots &\ddots &\vdots&\vdots\\
\overline{f}(j_{Q(N+1)},q_1|x,t) f(j_{Q(1)},q_1|y,t') &\cdots & \overline{f}(j_{Q(N+1)},q_N|x,t) f(j_{Q(N)},q_N|y,t') &
\overline{\phi}_{j_{Q(N+1)}}(x,t)\phi_{j_{Q(N+1)}}(y,t')
\end{array}
\right|
\end{align*}
In the previous result $q_i$ appears only in the $i$-th column which means that we can sum inside the determinant. Introducing two square
matrices of dimension $N+1$, depending on the state $\bj$, with elements
\begin{align}\label{defum}
\tilde{U}_{ab}^{(-)}(x,t;y,t')&=\sum_{q=1}^\infty \overline{f}(j_{a},q|x,t) f(j_{b},q|y,t')\, ,\ \ a,b=1,\cdots,N+1\, ,\\
\tilde{R}_{ab}^{(-)}(x,t;y,t')&=\overline{\phi}_{j_{a}}(x,t)\phi_{j_{b}}(y,t')\, ,\ \  a,b=1,\cdots,N+1\,
\end{align}
we obtain
\begin{align}
A_q&=\frac{1}{N!}\sum_{Q\in S_{N+1}}\left|
\begin{array}{cccc}
\tilde{U}^{(-)}_{Q(1),Q(1)} &\cdots & \tilde{U}^{(-)}_{Q(1),Q(N)}& \tilde{R}^{(-)}_{Q(1),Q(N+1)}\\
\vdots &\ddots &\vdots&\vdots\\
\tilde{U}^{(-)}_{Q(N+1),Q(1)} &\cdots & \tilde{U}^{(-)}_{Q(N+1),Q(N)}& \tilde{R}^{(-)}_{Q(N+1),Q(N+1)}
\end{array}
\right|(x,t;y,t')\, .
\end{align}
Performing permutations of both columns and rows such that $(Q(1),\cdots,Q(N+1))\rightarrow (1,\cdots,N+1)$ we find
\begin{align}
A_q&=\sum_{k=1}^{N+1}\left|
\begin{array}{cccc}
\tilde{U}^{(-)}_{1,1} &\cdots & \tilde{R}^{(-)}_{1,k}& \tilde{U}^{(-)}_{1,N+1}\\
\vdots &\ddots &\vdots&\vdots\\
\tilde{U}^{(-)}_{N+1,1} &\cdots & \tilde{R}^{(-)}_{N+1,k}& \tilde{U}^{(-)}_{N+1,N+1}
\end{array}
\right|(x,t;y,t')\, .
\end{align}
which can be written as
\begin{align}
A_q&=\frac{\6}{\6 z}\det_{N+1}\left. \left(\tilde{U}^{(-)}+z\tilde{R}^{(-)}\right)\right|_{z=0}\, ,\\
&=\det_{N+1}\left(\tilde{U}^{(-)}+\tilde{R}^{(-)}\right)-\det_{N+1}\tilde{U}^{(-)}\, ,
\end{align}
due to the fact that the matrix  $\tilde{R}^{(-)}$ has rank 1.

\subsubsection{Summation over $\mu_1,\cdots,\mu_M$}

We have obtained that
\begin{align}\label{aj5}
A&=\frac{1}{M!}\sum_{\mu_1}\cdots\sum_{\mu_M}\frac{e^{-i (t-t')\mu_\uparrow}}{(N+1)^M N^M} |\det_{M}B_\uparrow(\bl,\bmu)|^2\left[\det_{N+1}
\left(\tilde{U}^{(-)}+\tilde{R}^{(-)}\right)-\det_{N+1}\tilde{U}^{(-)}\right]\, ,
\end{align}
where the term appearing in the square parenthesis depends on $\mu_1,\cdots,\mu_M$ only via $e^{ i\Theta}=e^{i\sum_{a=1}^M\mu_a}$ [see (\ref{deff})
and (\ref{defum})]. This also means that the square parenthesis is periodic on $\Theta$ with period $2\pi$. Because $e^{i \mu_b N}=(-1)^{\overline{M}-1}$
or, equivalently, because $e^{i\Theta}$ are eigenvalues of the cyclic shift operator we have $\Theta=\frac{2\pi n}{N}$ with $n=0,1,\cdots,N-1$. Therefore,
in terms of the Kronecker symbol on $\mathbb{Z}_N$ defined by
\begin{align}
\delta_{(N)}(m)=\left\{
\begin{array}{ll}
1 &\mbox { if }   m=0\, (\mbox{mod } N)\, ,\\
0 & \mbox{ otherwise}\, ,
\end{array}
\right.
\, \ \ \
\delta_{(N)}(m)=\frac{1}{N}\sum_{p=0}^{N-1} e^{\frac{2\pi i}{N} pm }\, ,
\end{align}
a resolution of unity can be written as $ 1=\sum_{n=0}^{N-1}\delta_{(N)}\left (N\frac{\mu_1+\cdots+\mu_M}{2\pi}-n\right)\, .$
Defining $\tilde{U}^{(-)}_n= \tilde{U}^{(-)}|_{\Theta=2\pi n/N}$ (note that $R^{(-)}$ does not depend on $\Theta$) then (\ref{aj5}) can be written as
\begin{align}\label{aj6}
A&=\frac{1}{M!}\sum_{\mu_1}\cdots\sum_{\mu_M}\frac{e^{-i (t-t')\mu_\uparrow}}{(N+1)^M N^{M+1}} \sum_{n,p=0}^{N-1} e^{i p(\mu_1+\cdots+\mu_M)-\frac{2\pi i}{N}pn}
|\det_{M}B_\uparrow(\bl,\bmu)|^2\left[\det_{N+1}
\left(\tilde{U}^{(-)}_n+\tilde{R}^{(-)}\right)-\det_{N+1}\tilde{U}^{(-)}_n\right]\, ,
\end{align}
with  $[B_\uparrow(\bl,\bmu)]_{ab}=\sum_{n=1}^N e^{i n(\lambda_a-\mu_b)}$.
Let us focus on
\begin{align}
A_\mu=\frac{1}{M!}\sum_{\mu_1}\cdots\sum_{\mu_M}\frac{e^{i p(\mu_1+\cdots+\mu_M)}}{(N+1)^M N^{M+1}}
|\det_{M}B_\uparrow(\bl,\bmu)|^2\, .
\end{align}
Using the definition of the determinant we have
\begin{align}\label{aj7}
A_\mu&=\frac{1}{M!}\sum_{\mu_1}\cdots\sum_{\mu_M}\frac{e^{i p(\mu_1+\cdots+\mu_M)}}{(N+1)^M N^{M+1}}
\left(\sum_{P\in S_M}(-1)^P \prod_{a=1}^M \left[\overline{B}_\uparrow\right]_{P(a),a}\right)
\left(\sum_{Q\in S_M}(-1)^Q \prod_{a=1}^M \left[B_\uparrow\right]_{Q(a),a}\right)\, ,\nonumber\\
&=\frac{1}{M!}\sum_{\mu_1}\cdots\sum_{\mu_M}\frac{e^{i p(\mu_1+\cdots+\mu_M)}}{(N+1)^M N^{M+1}}
\sum_{Q\in S_M}\sum_{R\in S_M}(-1)^R \prod_{a=1}^M \left( \left[\overline{B}_\uparrow\right]_{RQ(a),a}\left[\overline{B}_\uparrow\right]_{Q(a),a}\right)\, ,\nonumber\\
&=\frac{1}{M!}\sum_{\mu_1}\cdots\sum_{\mu_M}\frac{e^{i p(\mu_1+\cdots+\mu_M)}}{(N+1)^M N^{M+1}}\sum_{Q\in S_M}
\left|
\begin{array}{ccc}
\left[\overline{B}_\uparrow\right]_{Q(1),1}&\cdots&\left[\overline{B}_\uparrow\right]_{Q(1),M}\\
\vdots &\ddots&\vdots\\
\left[\overline{B}_\uparrow\right]_{Q(M),1}&\cdots&\left[\overline{B}_\uparrow\right]_{Q(M),M}
\end{array}
\right|\prod_{a=1}^M \left(\left[\overline{B}_\uparrow\right]_{Q(a),a}\right)\, ,\nonumber\\
&=\frac{1}{M!}\sum_{\mu_1}\cdots\sum_{\mu_M}\sum_{Q\in S_M}
\left|
\begin{array}{ccc}
\frac{\left[\overline{B}_\uparrow\right]_{Q(1),1} \left[B_\uparrow\right]_{Q(1),1}e^{i p \mu_1}}{N(N+1)} &\cdots&
\frac{\left[\overline{B}_\uparrow\right]_{Q(1),M} \left[B_\uparrow\right]_{Q(M),M}e^{i p \mu_M}}{N(N+1)}\\
\vdots &\ddots&\vdots\\
\frac{\left[\overline{B}_\uparrow\right]_{Q(M),1} \left[B_\uparrow\right]_{Q(1),1}e^{i p \mu_1}}{N(N+1)} &\cdots&
\frac{\left[\overline{B}_\uparrow\right]_{Q(M),M} \left[B_\uparrow\right]_{Q(M),M}e^{i p \mu_M}}{N(N+1)}\\
\end{array}
\right|\, .
\end{align}
In the last determinant of (\ref{aj7}) $\mu_j$ appears only in the $j$-th column so we can sum inside the determinant. Introducing
a set of  matrices of dimension $M$ denoted by $O_p^{(-,\uparrow)}$ with elements
\be
[O_p^{(-,\uparrow)}]_{ab}=\frac{1}{N(N+1)}\sum_{n=1}^N\sum_{m=1}^N\sum_\mu e^{i(p+m-n)\mu+i n\lambda_a-i m\lambda_b}\, , \ \  a,b=1,\cdots,M\, ,
\ee
 we obtain
\begin{align}\label{aj8}
A_\mu&=\frac{1}{N}\frac{1}{M!}\sum_{Q\in S_M}
\left|
\begin{array}{ccc}
\left[\overline{O}^{(\uparrow,-)}_p \right]_{Q(1),Q(1)}  &\cdots& \left[\overline{O}^{(\uparrow,-)}_p \right]_{Q(M),Q(1)}\\
\vdots &\ddots&\vdots\\
\left[\overline{O}^{(\uparrow,-)}_p \right]_{Q(1),Q(M)}  &\cdots& \left[\overline{O}^{(\uparrow,-)}_p \right]_{Q(M),Q(M)}
\end{array}
\right|\, .
\end{align}
Permuting the rows and columns such that $(Q(1),\cdots,Q(M))\rightarrow(1,\cdots,M)$ we obtain $M!$ identical terms. Plugging (\ref{aj8}) in (\ref{aj6})
we finally obtain
\begin{align}
A&=e^{- i(t-t')\mu_\uparrow}\frac{1}{N}\sum_{n,p=0}^{N-1} e^{-\frac{2\pi i}{N} pn}\det_M O_p^{(-,\uparrow)}
\left[\det_{N+1}\left(\tilde{U}^{(-)}_n+\tilde{R}^{(-)}\right)-\det_{N+1}\tilde{U}^{(-)}_n\right]\, ,
\end{align}
which represents the finite size determinant representation for the mean value $A\equiv\langle
\Phi_{N+1,M}(\bj,\bl)|\Psi_\uparrow^\dagger(x,t)\Psi_\uparrow(y,t')|\Phi_{N+1,M}(\bj,\bl)\rangle$.

\subsection{Determinant representation for $\langle \Phi_{N,\bar{M}}(\bq,\bmu)|\Psi_\uparrow(x,t)\Psi_\uparrow^\dagger(y,t')|\Phi_{N,\bar{M}}(\bq,\bmu)\rangle$}

In the case of the other type of mean value of bilocal operators $B=\langle \Phi_{N,\bar{M}}(\bq,\bmu)|\Psi_\uparrow(x,t)\Psi_\uparrow^\dagger(y,t')|\Phi_{N,\bar{M}}(\bq,\bmu)\rangle$
 we have ($\bar{M}=M$ for $\sigma=\uparrow$)
\begin{align}\label{i7}
B &=\sum_{\substack{j_1<\cdots<j_{N+1}\\  \lambda_1<\cdots<\lambda_M}}
\mathcal{F}_{N,M}^{(\uparrow)}(\bj,\bl;\bq,\bmu|x,t)
\overline{\mathcal{F}}_{N,M}^{(\uparrow)}(\bj,\bl;\bq,\bmu|y,t')\, ,\nonumber\\
&=\sum_{\substack{j_1<\cdots<j_{N+1}\\  \lambda_1<\cdots<\lambda_M}}\frac{e^{i(t-t')\mu_\uparrow}}{N^M(N+1)^M}|\det_{M} B_\uparrow(\bl,\bmu)|^2
\det_{N+1}D(\bj,\bq|x,t)\overline{\det_{N+1} D(\bj,\bq|y,t')}\, .
\end{align}
Like in the previous case the summation over $j$'s is independent on the summation over $\lambda$'s and the summands are independently
symmetric in $j$'s and $\lambda$'s and vanish when two of them are equal. Therefore, the summation can be written as
\be
\sum_{\substack{j_1<\cdots<j_{N+1}\\  \lambda_1<\cdots<\lambda_M}}=\frac{1}{(N+1)!}\sum_{j_1=1}^\infty\cdots\sum_{j_{N+1}=1}^\infty\frac{1}{M!}
\sum_{\lambda_1}\cdots\sum_{\lambda_M}\, ,
\ee
where $\sum_\lambda h(\lambda)=\sum_{l=1}^{N+1} h(\lambda_l)$ with $\lambda_l=\frac{2\pi}{N+1}\left(-\frac{N+1}{2}-\frac{1+(-1)^{N-M+1}}{4}+l\right)$ for an arbitrary
function $h$.

\subsubsection{Summation over $\lambda_1,\cdots,\lambda_{N+1}$}

The summation over the $\lambda$'s in (\ref{i7}) can be written as
\begin{align}\label{i8}
B_j&=\frac{1}{(N+1)!}\sum_{j_1=1}^\infty\cdots\sum_{j_{N+1}=1}^\infty\det_{N+1}D(\bj,\bq|x,t)\overline{\det_{N+1} D(\bj,\bq|y,t')}\, ,\nonumber\\
&=\frac{1}{(N+1)!}\sum_{j_1=1}^\infty\cdots\sum_{j_{N+1}=1}^\infty\sum_{P,Q\in S_{N+1}}(-1)^{P+Q}
\left(\prod_{l=1}^N f(j_{P(l)},q_l|x,t)\overline{f}(j_{Q(l)},q_l|y,t')\right)\phi_{j_{P(N+1)}}(x,t)\overline{\phi}_{j_{Q(N+1)}}(y,t')\, ,\nonumber\\
&=\frac{1}{(N+1)!}\sum_{j_1=1}^\infty\cdots\sum_{j_{N+1}=1}^\infty\sum_{R,Q\in S_{N+1}}(-1)^{R}
\left(\prod_{l=1}^N f(j_{RQ(l)},q_l|x,t)\overline{f}(j_{Q(l)},q_l|y,t')\right)\phi_{j_{RQ(N+1)}}(x,t)\overline{\phi}_{j_{Q(N+1)}}(y,t')\, ,\nonumber\\
&=\frac{1}{(N+1)!}\sum_{j_1=1}^\infty\cdots\sum_{j_{N+1}=1}^\infty\sum_{Q\in S_{N+1}}
\left(\prod_{l=1}^N\overline{f}(j_{Q(l)},q_l|y,t')\right)\overline{\phi}_{j_{Q(N+1)}}(y,t')\nonumber\\
&\qquad\qquad\qquad\qquad\qquad\times
\left|
\begin{array}{cccc}
f(j_{Q(1)},q_1|x,t)  &\cdots & f(j_{Q(1)},q_N|x,t)& \phi_{j_{Q(1)}}(x,t)\\
\vdots &\ddots &\vdots&\vdots\\
f(j_{Q(N+1)},q_1|x,t)  &\cdots & f(j_{Q(N+1)},q_N|x,t)& \phi_{j_{Q(N+1)}}(x,t)\\
\end{array}
\right|\, .
\end{align}
Multiplying the $j$-th row of the last determinant with $\overline{f}(j_{Q(j)},q_j|y,t')$ and the $N+1$-th row with $\overline{\phi}_{j_{Q(N+1}}(y,t')$
we see that we have $j_{Q(l)}$ appearing only on the $l$-th row which means that we can sum inside the determinant. Introducing the $\bq$ dependent matrix
and vectors
\begin{align}
\tilde{U}^{(+)}_{ab}(x,t;y,t')&=\sum_{j=1}^\infty f(j,q_b|x,t)\overline{f}(j,q_a|y,t')\, ,\ \ a,b,=1,\cdots,N\, ,\\
\tilde{e}_a(x,t;y,t')&=\sum_{j=1}^\infty f(j,q_a|x,t)\overline{\phi}_j(y,t')\, ,\ \ a=1,\cdots,N\, ,\\
\tilde{\bar{e}}_a(x,t;y,t')&=\sum_{j=1}^\infty\overline{f}(j,q_a|y,t')\phi_j(x,t)\, ,\ \ a=1,\cdots,N\, ,
\end{align}
and
\be
g(x,t;y,t')=\sum_{j=1}^\infty\phi_j(x,t)\overline{\phi}_{j}(y,t')\, ,
\ee
then (\ref{i8}) can be written as (the summation over the $Q$ permutations  gives $(N+1)!$ identical terms)
\begin{align}\label{i9}
B_j&=\left|
\begin{array}{cccc}
\tilde{U}_{1,1}^{(+)}  &\cdots & \tilde{U}_{1,N}^{(+)}& \tilde{\bar{e}}_1\\
\vdots &\ddots &\vdots&\vdots\\
\tilde{U}_{N,1}^{(+)}  &\cdots & \tilde{U}_{N,N}^{(+)}& \tilde{\bar{e}}_N\\
\tilde{e}_1&\cdots&\tilde{e}_N& g
\end{array}
\right|(x,t;y,t')\, .
\end{align}
Introducing the  $\bq$ dependent matrix
\begin{align}
\tilde{R}^{(+)}_{ab}(x,t;y,t')=\tilde{\bar{e}}_a(x,t;y,t')\tilde{e}_b(x,t;y,t')\, ,\ \ a,b,=1,\cdots,N\, ,
\end{align}
and expanding on the last column of (\ref{i9}) we obtain
\begin{align}
B_j&=\left[g+\frac{\6}{\6 z}\right]\det_N\left(\tilde{U}^{(+)}-z \tilde{R}^{(+)}\right)\, ,\nonumber\\
&=\det_N\left(\tilde{U}^{(+)}-\tilde{R}^{(+)}\right)+(g-1)\det_N\tilde{U}^{(+)}\, .
\end{align}

\subsubsection{Summation over $\lambda_1,\cdots, \lambda_M$}

We have obtained that
\begin{align}
B&=\frac{1}{M!}\sum_{\lambda_1}\cdots \sum_{\lambda_M} \frac{e^{i(t-t')\mu_\uparrow}}{N^M(N+1)^M}|\det_{M} B_\uparrow(\bl,\bmu)|^2
\left[\det_N\left(\tilde{U}^{(+)}-\tilde{R}^{(+)}\right)+(g-1)\det_N\tilde{U}^{(+)}\right]\, ,
\end{align}
where $\tilde{U}^{(+)}$ and $\tilde{R}^{(+)}$ depend on $\lambda_1,\cdots,\lambda_M$ only via their sum $\Lambda=\lambda_1+\cdots\lambda_M$.
Like in the previous case this implies periodicity in $\Lambda$ with period $2\pi$. In this case $\Lambda=\frac{2\pi n}{N+1}$ with $n=0,\cdots, N$
which means that a resolution of identity is given by $ 1=\sum_{m=0}^{N}\delta_{(N+1)}\left (m-(N+1)\frac{\la_1+\cdots+\la_M}{2\pi}\right)\, $
with
\begin{align}
\delta_{(N+1)}(m)=\left\{
\begin{array}{ll}
1 &\mbox { if }   m=0\, (\mbox{mod } N+1)\, ,\\
0 & \mbox{ otherwise}\, ,
\end{array}
\right.
\, \ \ \
\delta_{(N+1)}(m)=\frac{1}{N}\sum_{r=0}^{N} e^{\frac{2\pi i}{N} r m }\, ,
\end{align}
Introducing $\tilde{U}^{(+)}_m=\tilde{U}^{(+)}|_{\Lambda=2\pi m/(N+1)}$
$\tilde{R}^{(+)}_m=\tilde{R}^{(+)}|_{\Lambda=2\pi m/(N+1)}$ the computations are similar with the ones in the previous section
and \cite{IP98} obtaining
\begin{align}
B=e^{i (t-t')\mu_\uparrow}\frac{1}{N+1}\sum_{r,m=0}^Ne^{\frac{2\pi i}{N+1} rm}\det_M O_r^{(+,\uparrow)}
\left[\det_N\left(\tilde{U}^{(+)}_m-\tilde{R}^{(+)}_m\right)+(g-1)\det_N\tilde{U}^{(+)}_m\right]\, ,
\end{align}
with the $ O_r^{(+,\uparrow)} $  matrices defined as (we correct a typo in 4.44 of \cite{IP98})
\be
[O_r^{(+,\uparrow)}]_{ab}=\frac{1}{N(N+1)}\sum_{m=1}^N\sum_{n=1}^N\sum_\lambda e^{-i(r+m-n)\lambda-i n\mu_a+i m\mu_b}\, ,\ \ a,b=1,\cdots,M\, .
\ee

\subsection{Thermodynamic limit}

The thermal summation in (\ref{defgm}) and (\ref{defgp}) is very similar with the one performed in \cite{IP98} for two-component systems without an
external potential (see also \cite{P22b}). The main ingredient is the von Koch determinant formula which reads
\begin{align}\label{vonK}
\det(1+z A)&=1+z\sum_{m=1}^M A_{m,m}+\frac{z^2}{2!}\sum_{m=1}^M\sum_{n=1}^M
\left|
\begin{array}{cc}
A_{m,m}& A_{m,n}\\
A_{n,m}& A_{n,n}
\end{array}
\right|+\cdots
\end{align}
for $A$  a square matrix of dimension $M$ (which can also be infinite) and $z$ a bounded complex parameter. Following the similar steps in \cite{IP98}
one obtains (\ref{gm}) and (\ref{gp}).

\section{Thermodynamics of the impenetrable Gaudin-Yang model}\label{a1}

In this Appendix we present some results for the thermodynamics of the trapped impenetrable Gaudin-Yang model. The energy spectrum
of the trapped impenetrable system is given by (\ref{energy}). We notice two important features: a) it is independent of the statistics
of the constituent particles and b) does not depend on the spin state $\bl$. This means that for the system with $N$ particles of which
$M$ have spin down there are $C^N_M$ states with the same energy for a given set of orbital numbers $\bq$. The partition function is
\begin{align}\label{z}
\mathcal{Z}(\mu,B,T)&=\mbox{Tr}\left[e^{-H_I/T}\right] =\sum_{N=0}^\infty\sum_{M=0}^N \sum_{q_1<\cdots<q_N}\sum_{\mu_1<\cdots<\mu_M}e^{-E_{N,M}(\bq)/T}\, ,\nonumber\\
&=\sum_{N=0}^\infty\sum_{M=0}^N \sum_{q_1<\cdots<q_N}\sum_{\mu_1<\cdots<\mu_M}e^{2 BM/T}e^{-\sum_{i=1}^N(\varepsilon(q_i)-\mu+B)/T}\, ,\nonumber\\
&=\sum_{N=0}^\infty \sum_{q_1<\cdots<q_N}\left(1+e^{2 B/T}\right)^Ne^{-\sum_{i=1}^N(\varepsilon(q_i)-\mu+B)/T}\, ,\nonumber\\
&=\sum_{N=0}^\infty \sum_{q_1<\cdots<q_N}\left(2\cosh(B/T)\right)^Ne^{-\sum_{i=1}^N(\varepsilon(q_i)-\mu)/T}\, ,\nonumber\\
&=\prod_{q=1}^\infty \left(1+2\cosh(B/T)e^{-(\varepsilon(q)-\mu)/T}\right)
\end{align}
where we have used $\sum_{M=0}^N \sum_{\mu_1<\cdots<\mu_M}e^{2 BM/T}=\sum_{M=0}^N C^N_M e^{2 B/T}=\left(1+e^{2 B/T}\right)^N$. The grandcanonical
potential $\phi(\mu,B,T)=U-TS-\mu(N_\uparrow+N_\downarrow)+B(N_\uparrow-N_\downarrow)$  is
\begin{align}
\phi(\mu,B,T)=-T\ln \mathcal{Z}(\mu,B,T)=-T\sum_{q=1}^\infty\ln\left(1+2\cosh(B/T)e^{-(\varepsilon(q)-\mu)/T}\right)\, .
\end{align}
From the grandcanonical potential the number of particles of each type can be obtained as
\begin{align}
N_{\uparrow} & =-\frac{1}{2}\left(\frac{\6\phi}{\6\mu}-\frac{\6\phi}{\6B}\right)=\sum_{q=1}^\infty\frac{e^{-B/T}}{2\cosh(B/T)+e^{(\varepsilon(q)-\mu)/T}}\, , \\
N_\downarrow & =-\frac{1}{2}\left(\frac{\6\phi}{\6\mu}+\frac{\6\phi}{\6B}\right)=\sum_{q=1}^\infty\frac{e^{B/T}}{2\cosh(B/T)+e^{(\varepsilon(q)-\mu)/T}}\, .
\end{align}

\section{Elements of the $V^{(T,-)}$ matrix in the equal-time case}\label{a2}

Here we derive the simplified expressions for the elements of the $V^{(T,-)}$ matrix in the equal-time case (\ref{i10}).
Because  $[V^{(T,-)}]_{ab}=\sqrt{\vartheta(a)}(U_{ab}^{(-)}-\delta_{a,b})\sqrt{\vartheta(b)}$   we will focus on  $U^{(-)}_{ab}$  which is
defined in (\ref{defumT}). We obtain different results depending on the ordering of $x$ and $y$. In the case $x\le y$ using (\ref{defft})
we have
\begin{align}\label{ai1}
[U^{(-)}]_{ab}&=\delta_{a,b}-\zeta \int_{y}^{L_+} \overline{\phi}_a(w,t)\phi_b(w,t)\, dw-\overline{\zeta}\int_{x}^{L_+}\overline{\phi}_a(v,t)\phi_b(v,t)\, dv\nonumber\\
&\qquad\qquad +\zeta\overline{\zeta}\sum_{q=1}^\infty\left(\int_{x}^{L_+}\overline{\phi}_a(v,t)\phi_q(v,t)\, dv\right)
\left(\int_{y}^{L_+}\overline{\phi}_q(w,t)\phi_b(w,t)\, dw\right)\, ,\nonumber\\
&=\delta_{a,b}-\overline{\zeta}\int_{x}^{y}\overline{\phi}_a(v,t)\phi_b(v,t)\, dv -
\underbrace{(\zeta+\overline{\zeta})\int_{y}^{L_+} \overline{\phi}_a(w,t)\phi_b(w,t)\, dw}_{A}\nonumber\\
&\qquad\qquad +\underbrace{\zeta\overline{\zeta}\sum_{q=1}^\infty\left(\int_{x}^{y}\overline{\phi}_a(v,t)\phi_q(v,t)\, dv\right)
\left(\int_{y}^{L_+}\overline{\phi}_q(w,t)\phi_b(w,t)\, dw\right)}_{B}\nonumber\\
&\qquad\qquad +\underbrace{\zeta\overline{\zeta}\sum_{q=1}^\infty\left(\int_{y}^{L_+}\overline{\phi}_a(v,t)\phi_q(v,t)\, dv\right)
\left(\int_{y}^{L_+}\overline{\phi}_q(w,t)\phi_b(w,t)\, dw\right)}_{C}\, ,
\end{align}
with $\zeta=(1-e^{i(\pi\kappa-\eta)})$. Now we will show that in the previous expression the terms $A$ and $C$ are equal cancelling each other.
We have
\begin{align}\label{ai2}
\int_{y}^{L_+} \overline{\phi}_a(w,t)\phi_b(w,t)\, dw=\int_{L_-}^{L_+}\overline{\tilde{\phi}}_a(w,t)\tilde{\phi}_b(w,t)\, dw\,
\end{align}
where $\tilde{\phi}_a(w,t)=\bm{1}_{[y,L_+]}\phi_a(w,t)$ and $\tilde{\phi}_b(w,t)=\bm{1}_{[y,L_+]}\phi_b(w,t)$ with $\bm{1}_{[y,L_+]}$
the characteristic function of the interval $[y,L_+]$ which is $1$ when $w$ is in the interval and $0$ otherwise. Also, using the
orthonormality of the wavefunctions (\ref{ortho}) we find
\begin{align}\label{ai3}
\int_{L_-}^{L_+}\overline{\tilde{\phi}}_a(w,t)\tilde{\phi}_b(w,t)\, dw&=
\int_{L_-}^{L_+}\int_{L_-}^{L_+}\overline{\tilde{\phi}}_a(v,t)\delta(v-w)\tilde{\phi}_b(w,t)\, dw dv\, ,\nonumber\\
&=\sum_{q=1}^\infty\left(\int_{L_-}^{L_+}\overline{\tilde{\phi}}_a(v,t)\phi_q(v,t)\, dv\right)
\left(\int_{L_-}^{L_+}\tilde{\phi}_q(w,t)\tilde{\phi}_b(w,t)\, dw\right)\, ,\nonumber\\
&=\sum_{q=1}^\infty\left(\int_{y}^{L_+}\overline{\phi}_a(v,t)\phi_q(v,t)\, dv\right)
\left(\int_{y}^{L_+}\overline{\phi}_q(w,t)\phi_b(w,t)\, dw\right)\, .
\end{align}
Eqs.~(\ref{ai2}) and (\ref{ai3}) together with $\zeta+\overline{\zeta}=\zeta\overline{\zeta}=2-2\cos(\pi\kappa-\eta)$ show that the
$A$ and $C$ terms cancel each other in (\ref{ai1}). In a similar fashion it can be shown that $B=0$ by noticing that it is the
expansion of $\int_{L_-}^{L_+}\overline{\tilde{\phi}}_a(w,t)\tilde{\phi}_b(w,t)\, dw$ with $\tilde{\phi}_a(w,t)=\bm{1}_{[x,y]}\phi_a(w,t)$ and $\tilde{\phi}_b(w,t)=\bm{1}_{[y,L_+]}\phi_b(w,t)$ and $\bm{1}_{[x,y]}\bm{1}_{[y,L_+]}=0$. Therefore, we find
\be
[V^{(T,-)}]_{ab}=-\left(1-e^{-i[\pi\kappa-\eta]}\right)\sqrt{\vartheta(a)\vartheta(b)}
\int_x^y\overline{\phi}_a(v,t)\phi_b(v,t)\, dv\, , \ \  x\le y\, .
\ee
In the other case we obtain
\be
[V^{(T,-)}]_{ab}=-\left(1-e^{+i[\pi\kappa-\eta]}\right)\sqrt{\vartheta(a)\vartheta(b)}
\int_y^x\overline{\phi}_a(v,t)\phi_b(v,t)\, dv\, , \ \  y<x\, .
\ee

\section{Equivalence with Lenard's formula}\label{a3}

We will show the equivalence of the determinant representation (\ref{statictfull}) with Lenard's formula (\ref{lenard}). Similar with
the  particular case of zero temperature treated in Sec.\ref{s5} the representation (\ref{lenard}) can be understood as the first Fredholm minor of the
integral operator $1-\xi\, \hat{g}_\uparrow^{FF}$ acting on $[x,y]$  with kernel $g_\uparrow^{FF}(x,y|\,t)=\sum_{a=1}^\infty
\vartheta(a)\, \overline{\phi}_a(x,t)\phi_a(y,t)$ and $\xi$ defined in $(\ref{defxi})$. From Hurwitz formula \cite{H14} we have
\be\label{a40}
g_\uparrow^{(-)}(x,y|\, t)=R_\uparrow^{FF}(x,y|\, t)\det\left(1-\xi\, \hat{g}_\uparrow^{FF}\right)\, ,
\ee
with the resolvent satisfying the integral equation
\be\label{a41}
R_\uparrow^{FF}(\lambda,\mu|\, t)=g_\uparrow^{FF}(\lambda,\mu|\, t)+\xi\int_x^yg_\uparrow^{FF}(\lambda,\nu|\, t)R_\uparrow^{FF}
(\nu,\mu|\, t)\, d\nu\, .
\ee
We will show that (\ref{statictfull}) is equivalent with (\ref{a40}) but, first we need a preliminary result. For any invertible matrix
$A$  and two column vectors of the same dimension, $u$ and $v$, the following identity holds: $\det(A+uv^T)=\det A+\det A v^T A^{-1} u$
\cite{M90}. Introducing  $\phi_a^T(x,t)=\sqrt{\vartheta(a)}\phi_a(x,t)$ and noticing that the matrix $r^{(T,-)}$ defined in (\ref{defrt}) can be
written as $u v^T$ with $u=(\overline{\phi}_1^T(x,t),\overline{\phi}_2^T(x,t),\cdots)^T$ and $v=(\phi_1^T(y,t),\phi_2^T(y,t),\cdots)^T$
we find from (\ref{statictfull})
\be\label{ai11}
g_\uparrow^{(-)}(x,y|\,t)=\det(1+v^{(T,-)})\sum_{i,j} \phi_i^T(y,t)\left[\left(1+v^{(T,-)}\right)^{-1}\right]_{ij} \overline{\phi}_j^{T}(x,t)\, .
\ee
The proof that $\det(1+v^{(T,-)})=\det\left(1-\xi \hat{g}_\uparrow^{FF}\right)$ is the same as in Sec.~V.B of \cite{P20}. It remains to
show that the other term in the right hand side of (\ref{ai11}) is equal to $R_\uparrow^{FF}(x,y|\, t)$. In terms of $\phi_a^T$ we have
$g_\uparrow^{FF}(\lambda,\mu|\, t)=\sum_{a=1}^\infty\overline{\phi}_a^T(\lambda,t)\phi_a^T(\mu,t)$. Plugging this in the equation for the
resolvent (\ref{a41})
we find
\be\label{ai11b}
R_\uparrow^{FF}(\lambda,\mu|\, t)=\sum_{b=1}^\infty \overline{\phi}_a^T(\lambda,t)\phi_a^T(\mu,t)+\xi\sum_{b=1}^\infty\overline{\phi}^T_b(\lambda,t) B_b(\mu,t)\, ,
\ee
with $B_b(\mu,t)=\int_x^y\phi_b^T(\nu,t)R_\uparrow^{FF}(\nu,\mu|\, t)\, d\nu$. In order to obtain the $B_b(\mu,t)$ coefficients we multiply the previous
expression with $\phi_a^T(\lambda,t)$ and integrate from $x$ to $y$. We obtain
\be\label{ai12}
B_a(\mu,t)=\sum_{b=1}^\infty A_{ba}(t)\phi_b^T(\mu,t)+\xi\sum_{b=1}^\infty A_{ba}(t)B_b(\mu,t)\, ,
\ee
where we have introduced the matrix $A$ with elements
\be\label{defa}
A_{ab}(t)=\int_x^y \overline{\phi}_a^T(\lambda,t)\phi_b^T(\lambda,t)\, d\lambda\, .
\ee
In terms of the column vectors $\boldsymbol{\phi}=(\phi_1^T(\mu,t),\phi_2^T(\mu,t),\cdots)^T$, $B=(B_1(\mu,t),B_2(\mu,t),\cdots)^T$ the equation (\ref{ai12})
 can be written as $B=A^T\boldsymbol{\phi}+\xi A^T B$ with the solution $B=\left(1-\xi A^T\right)^{-1}A^T\boldsymbol{\phi}$. Using this result and (\ref{ai11b})
 we have $R_\uparrow^{FF}(\lambda,\mu|\, t)=\overline{\boldsymbol{\phi}}^T\left(1+\xi \left(1-\xi A^T\right)^{-1}A^T\right)\boldsymbol{\phi}$ with
 $\overline{\boldsymbol{\phi}}^T=(\overline{\phi}_1^T(\lambda,t),\overline{\phi}_2^T(\lambda,t),\cdots)$ a row vector. This last relation can also be written as
 $R_\uparrow^{FF}(\lambda,\mu|\, t)=\overline{\boldsymbol{\phi}}^T \left(1-\xi A^T\right)^{-1}\boldsymbol{\phi}$ and shows that
 \be\label{ai13}
 R_\uparrow^{FF}(x,y|\, t)= \sum_{i,j} \overline{\phi}_i^T(x,t)\left[ \left(1-\xi A^T\right)^{-1}\right]_{ij} \phi_j^{T}(y,t)\, .
 \ee
 Using $\left(1-\xi A^T\right)^{-1}=\left[\left(1-\xi A\right)^{-1}\right]^T$ this shows that (\ref{ai13}) is equal to the second term in the right hand side of (\ref{ai11})
 proving the equivalence of the representations (\ref{lenard}) and (\ref{statictfull}).

\end{document}